\documentclass[10pt,citeautoscript,floatfix,aps,prb,twocolumn,
      superscriptaddress,longbibliography]{revtex4-2}

\usepackage{amsmath}
\usepackage{physics}
\usepackage{graphicx}
\usepackage{epstopdf}
\usepackage{natbib}
\usepackage{multirow}
\usepackage{array}
\usepackage[usenames,dvipsnames]{color}
\graphicspath{{figs/}}
\usepackage{dcolumn}
\usepackage{bm}
\usepackage{soul}  
\usepackage{changes}
\def\add#1{\added{#1}}

\newcommand{\e}{\epsilon}
\newcommand{\w}{\omega}

\newcommand{\eq}[1]{Eq.~(\ref{#1})}
\newcommand{\loo}{LiOsO$_3$}
\newcommand{\sto}{SrTiO$_3$}

\newcommand{\SR}{{\text{SR}}}

\newcommand{\PhiLR}{$\Phi^{\text{LR}}$}
\newcommand{\PhiSR}{$\Phi^{\text{SR}}$}
\newcommand{\Vloc}{$V^{\text{loc}}$}
\newcommand{\Velec}{$V^{\text{pc}}$}
\newcommand{\Zstat}{${\bf Z}^{\rm stat}_{\kappa\alpha}$}
\newcommand{\Zstatcub}{$Z^{\rm stat}_\kappa$}
\newcommand{\Zdyncub}{$Z^{\rm dyn}_\kappa$}

\begin{document}
\title{Static Born charges and quantum capacitance in metals and doped semiconductors}
\author{Asier Zabalo}
\affiliation{Institut de Ci\`{e}ncia de Materials de Barcelona (ICMAB-CSIC), Campus UAB, 08193 Bellaterra, Spain}
\author{Cyrus E. Dreyer}
\affiliation{Department of Physics and Astronomy, Stony Brook University, Stony Brook, New York, 11794-3800,
  USA}
\affiliation{Center for Computational Quantum Physics, Flatiron Institute, 162 5th Avenue, New York, New York 10010,
USA}
\author{Massimiliano Stengel}
\affiliation{Institut de Ci\`{e}ncia de Materials de Barcelona (ICMAB-CSIC), Campus UAB, 08193 Bellaterra, Spain}
\affiliation{ICREA-Instituci\'{o} Catalana de Recerca i Estudis Avan\c{c}ats, 08010 Barcelona, Spain}
\date{\today}
\begin{abstract}
Born dynamical charges ($\textbf{Z}^{\rm dyn}$) play a key role in the lattice dynamics of most crystals, including both insulators and  
metals in the nonadiabatic (``clean'') regime. 
Very recently, the so-called static Born charges, $\textbf{Z}^\text{stat}$, were introduced [G. Marchese, \textit{et al.}, Nat. Phys. \textbf{20}, 88 (2024)] as a means to modeling the long-wavelength behavior of polar phonons in overdamped (``dirty'') metals. 
Here we present a method to calculate $\textbf{Z}^\text{stat}$ directly at the zone center, by applying the ``$2n+1$'' theorem to the long-wavelength expansion of the charge response to a phonon. Furthermore, we relate $\textbf{Z}^\text{stat}$ to the charge response to a uniform strain perturbation via an exact sum rule, where the quantum capacitance of the material plays a crucial role.
We showcase our findings via extensive numerical tests on simple metals aluminum and copper, polar metal \loo{}, and doped semiconductor \sto{}.
Based on our results, we critically discuss the physical significance of  
$\textbf{Z}^\text{stat}$ in light of their
dependence on the choice of the electrostatic reference, and on the length scale that is assumed in the definition of the macroscopic potentials.
\end{abstract}

\maketitle

\section{Introduction}

It has been well known for the better part of a century that long-range electrostatic interactions play a key role in the lattice dynamics of insulating systems \cite{Lyddane1941,Born1954,Cochran1962}. The fundamental theory thereof, established by Pick, Cohen and Martin in 1970 \cite{Pick1970}, paved the way to modern \textit{ab-initio} implementations \cite{GonzeLee1997,Baroni2001,Verdi2015,Sjakste2015,Giustino2017} of phonons and electron-phonon interactions in real materials. 
Recent works have generalized this idea to doped semiconductors and metals \cite{Bistoni2019,Binci2020,Dreyer2022,Wang2022}, demonstrating that in certain regimes, long-range electrostatic interactions related to phonons can occur in metals due to nonadiabatic effects. Specifically, this occurs for the ``clean limit'' of a metal or doped semiconductor, i.e., under the assumption that an optical phonon of frequency $\omega$ is much larger that
the inverse carrier lifetime, $1/\tau$
\cite{Calandra2010,Saitta2008}. This physics is encapsulated in the ``nonadiabatic'' Born dynamical charges  
\cite{Saitta2008,Bistoni2019,Binci2020,Dreyer2022}, whose sublattice sum relates to the inertia of the free carriers in the material.

At the level of the static force constants, of direct relevance to ``dirty metals'' where $\w \ll 1/\tau$ \cite{Marchese2023}, no such long-range interactions exist, and (assuming a finite electronic temperature) the dynamical matrix is an analytic function of the momentum~\footnote{At low temperatures, discontinuities in the Fermi occupation function leads to the so-called Kohn anomalies, whose description requires a materials-specific discussion; they are removed at any finite electronic temperature (or smearing), which we assume in this work.}. 
Yet, even in this regime there are good 
reasons to seek a careful treatment of the 
long-wavelength limit.
For a weakly doped semiconductor, for example,  the free carriers have significant impact on the calculated phonon frequencies only in a narrow region of the Brillouin zone
surrounding the $\Gamma$ point \cite{Macheda2022}. 
This can result in an abrupt transition between an insulating regime, with a sizable frequency splitting between longitudinal (LO) and transverse (TO) optical branches, to a metallic one over a very short distance in momentum space.
Such a sharp feature would be clearly problematic to treat via
the established numerical interpolation algorithms.

The question then is what definition of the Born effective charge (BEC) should be used in modeling the long-wavelength limit of polar phonons \cite{GonzeLee1997,Baroni2001,Verdi2015,Sjakste2015,Giustino2017} in damped metals. 
Neither the macroscopic polarization nor its parametric dependence on the atomic positions can be defined or calculated in conductors, ruling out a straightforward approach.
To address this issue, a static version of the Born effective charges, \Zstat{} ($\kappa$ denotes the sublattice and $\alpha$ the Cartesian direction), was recently proposed \cite{Marchese2023} via ``macroscopic unscreening'' (i.e., with long-range electrostatic effects suppressed) of the charge response to the lattice distortion.
The \Zstat{} were found to be useful tools in the calculation  of electronic transport, optical properties of metals, and phonon-mediated superconductivity \cite{Marchese2023}.
Whether the \Zstat{} are
experimentally measurable quantities on their own,
however, still remains to be clarified.
Indeed, a 
direct measurement of the static BECs (e.g., via infrared absorption) appears unfeasible, as the LO/TO splitting vanishes in damped metals. 
To devise alternative experimental routes, establishing exact formal relations between the \Zstat{} and other observables is an obvious prerequisite.
In these regards, a sum rule for the static BECs in metals was demonstrated in Ref.~\citenum{Marchese2023}, but its physical significance was not clarified.
This situation contrasts that of
the 
Born dynamical charges, where an exact sum rule links their sublattice sum to a defining geometric property of the electronic ground state, 
the Drude weight tensor \cite{Dreyer2022}. 

From the methodological perspective there is considerable room for progress too. All the calculations of the static charges that were reported so far were based on  numerical differentiation of momentum-dependent response functions (either the charge response to a phonon \cite{Marchese2023} or the atomic force response to a modulated scalar potential \cite{Macheda2024}). 
The finite-difference approach to the long-wavelength expansion, while efficient and straightforward to implement, introduces additional parameters in the calculation, which must be carefully checked for convergence.
For many applications, it would be desirable to have a more straightforward code implementation, allowing for the calculation of the static charges directly at the $\Gamma$ point of the Brillouin zone.
The recent generalization of long-wavelength perturbation theory to metals~\cite{zabalo2024ensemble} appears ideally suited to this task.

In this work, we study the static BECs in three-dimensional (3D) metals with the aim of addressing these questions. First, we use the
formalism of Ref.~\cite{zabalo2024ensemble} to establish a closed formula for the static BECs, only requiring the calculation of response functions at the $\Gamma$ ($q=0$) point in the Brillouin zone. Next, we establish an exact sum rule for the static BECs, by relating them to the first-order charge response to a uniform strain. The latter, in turn, can be written as a deformation potential times the \emph{quantum capacitance} $C_Q$ of the 3D crystal, describing the number of electrons that are stored in the unit cell upon a shift of the Fermi level. Similarly, we find that the quantum capacitance relates the static BECs with the octupolar moment of the screened charge response to an atomic displacement. 
We demonstrate these formal results with \textit{ab-initio} density-functional perturbation theory (DFPT) calculations performed on aluminum, copper, \loo{}, and doped \sto{}.
Finally, we discuss the dependence of both \Zstat{} and $C_Q$ on the choice of the electrostatic reference, on the pseudopotential approximation, and on the specifics of how the electrostatic kernel is partitioned between an analytic and a nonanalytic part.

The rest of the paper is organized as follows. 
In Sec. \ref{sec:form}, we give our formal results, beginning with some basic definitions and a summary of the well-established strategies for addressing long-range electrostatic interactions in insulators; we then turn to metals, where we discuss the separation between short- and long-range interactions, the relation to quantum capacitance, and the derivation of a formal sum rule for $\mathbf{Z}^\text{stat}$. 
In Sec. \ref{sec:comp}, we describe the computational approach employed to access 
$\mathbf{Z}^\text{stat}$, which is based on the recently developed long-wavelength DFPT for metals.
In Sec. \ref{sec:res}, we present our results for $\mathbf{Z}^\text{stat}$ and the quantum capacitance of the selected materials. 
Finally, in Sec. \ref{sec:conc}, we conclude the paper with a summary of our findings and the conclusions.

\section{Theory \label{sec:form}}
Though much of the discussion in this section is more general, we will perform our derivations in the context of Kohn-Sham (KS) density-functional theory (DFT) \cite{Kohn1965}. 
Thus we will assume a set of single-particle Bloch wave functions $\psi_{n\textbf{k}}(\textbf{r})=u_{n\textbf{k}}(\textbf{r})e^{i\textbf{k}\cdot\textbf{r}}$, where $u_{n\textbf{k}}(\textbf{r})$ is cell-periodic and $n$ and \textbf{k} are the band and $k$-point indices, respectively. 
These define the ground state of the crystal, and are determined from a self-consistent solution of the KS Hamiltonian. The latter includes a single-particle kinetic term, an external potential (or pseudopotential) that describes the interaction with the nuclei, and density-dependent Hartree and exchange-correlation (XC) terms, which capture the electron-electron interaction. 
\subsection{Basic definitions}
A phonon perturbation is conveniently represented as a displacement of sublattice $\kappa$ in direction $\alpha$ modulated by a wavevector \textbf{q},
\begin{equation}
\Delta \tau^l_{\kappa\alpha}=\tau^{\textbf{q}}_{\kappa\alpha}e^{i\textbf{q}\cdot\textbf{R}_l},
\end{equation}
where $l$ labels a unit cell in the crystal, and $\textbf{R}_l$ is the lattice vector of that unit cell. 
The main quantity that we shall be dealing with is the 
force-constant matrix in reciprocal space, defined as the second derivative of the
energy with respect to atomic displacements \cite{GonzeLee1997}, 
\begin{equation}\label{Eq_FC}
\Phi_{\kappa\alpha,\kappa'\beta}(\textbf{q})=\frac{\partial^2 E_{\text{tot}}}{\partial \tau_{\kappa\alpha}^{-\textbf{q}} \partial\tau_{\kappa'\beta}^{\textbf{q}}}.
\end{equation}
We shall also consider an external electrostatic potential perturbation, where the variation in the external potential acting on the electrons takes the following form,  \cite{Royo2021}
\begin{equation}
    \Delta V^{\text{ext}}(\textbf{r})=-\varphi^{\textbf{q}}e^{i\textbf{q}\cdot\textbf{r}}.
\end{equation}
(The minus sign accounts for the negative electron charge.)
This allows us to write the 
macroscopic charge-density response to a phonon as  \cite{Royo2021}
\begin{equation}
\rho_{\kappa\alpha}(\textbf{q})=\frac{1}{\Omega}\frac{\partial^2 E_{\text{tot}}}{\partial \varphi^{-\textbf{q}} \partial\tau_{\kappa\alpha}^{\textbf{q}}},
\end{equation}
which can be equivalently written as cell average of the microscopic charge response to a phonon, $\rho_{\kappa \alpha}^{\bf q}({\bf r})$, 
\begin{equation}\label{Eq_rho_q}
\rho_{\kappa \alpha}({\bf q})= \frac{1}{\Omega} \int_\Omega d^3 r \, \rho^{\bf q}_{\kappa\alpha}(\bf r),
\end{equation}
where $\Omega$ is the volume of the unit cell.
The latter, in turn, can be written as 
\begin{equation}
\label{eq:micro_rho}
\begin{split}
\rho_{\kappa \alpha}^{\bf q}({\bf r}) =& 
Z_\kappa \sum_l\left(-\frac{\partial\delta(\mathbf{r}-\mathbf{R}_{l\kappa})}{\partial r_\alpha} -iq_\alpha \delta(\mathbf{r}-\mathbf{R}_{l\kappa})\right) \\
& -\int_\text{BZ} [d^3 k] \sum_{ij} \frac{f_{j\bf k+q}-f_{i\bf k}}{\e_{j\bf k+q}-\e_{i\bf k}} \times
\\
&\qquad  \langle u_{i \bf k}|{\bf r} \rangle \langle {\bf r}| u_{j \bf k+q} \rangle 
\langle u_{j \bf k+q}| \hat{\mathcal{H}}^{\tau_{\kappa \alpha}}_{\bf k,q} |u_{i \bf k} \rangle.
\end{split}
\end{equation}
The first term on the right-hand side embodies the contribution of the ionic point charges,~\cite{Stengel2013} of magnitude $Z_\kappa$,
with 
$\mathbf{R}_{l\kappa}=\mathbf{R}_l+\boldsymbol{\tau}_\kappa$, where
$\boldsymbol{\tau}_\kappa$ indicates the position 
of ion $\kappa$
within the unit cell.
In the second term, of electronic origin, we have used the shorthand notation $[d^3k]=\Omega/(2\pi)^3 d^3k$ for the 
Brillouin Zone (BZ) integration.
In Eq.~(\ref{eq:micro_rho}), 
$\hat{\mathcal{H}}^{\tau_{\kappa \alpha}}_{\bf k,q}=e^{-i(\textbf{k}+\textbf{q})\cdot\textbf{r}}\hat{\mathcal{H}}^{\tau_{\kappa \alpha}}e^{i\textbf{k}\cdot\textbf{r}}$ 
is the cell-periodic part of the first-order Hamiltonian with respect to a phonon perturbation. 
The calligraphic symbol $\hat{\mathcal{H}}^\lambda$ indicates that self-consistent (SCF) fields are included via the first-order density in the Hartree and exchange-correlation (XC) contributions; $H^\lambda$ will be used instead to indicate the ``external potential'' at first order in $\lambda$ (i.e., without SCF potentials).

For what follows, we will also need the second derivative with respect to $\varphi$, which defines
the so-called macroscopic reducible (or screened) susceptibility,
\begin{equation}
\label{eq:chi}
\chi(\textbf{q})=\frac{1}{\Omega}\frac{\partial^2 E_{\text{tot}}}{\partial \varphi^{-\textbf{q}} \partial\varphi^{\textbf{q}}},
\end{equation}
Similarly to the phonon case, this can also be written as an average over the primitive cell volume of the microscopic density response to a scalar potential perturbation,
\begin{equation}
\label{eq:chi_q}
\chi(\textbf{q})= \frac{1}{\Omega} \int_\Omega d^3 r \, \rho^{\bf q}_{\varphi}(\bf r).
\end{equation}
where
\begin{equation}
\label{Eq_capacitance}
\begin{split}
\rho_{\varphi}^{\bf q}({\bf r}) &= -\int_\text{BZ} [d^3 k] \sum_{ij} \frac{f_{i\bf k+q}-f_{j\bf k}}{\e_{i\bf k+q}-\e_{j\bf k}} \times
\\
&\langle u_{i \bf k}|{\bf r} \rangle \langle {\bf r}| u_{j \bf k+q} \rangle 
\langle u_{j \bf k+q}| \hat{\mathcal{H}}^{\varphi}_{\bf q} |u_{i \bf k} \rangle.
\end{split}
\end{equation}
Note that if SCF are neglected in calculating Eq.~(\ref{eq:chi}), then $\chi(\textbf{q})$ corresponds to the independent-particle susceptibility $\chi_0(\textbf{q})$.
\subsection{Long-range electrostatic interactions in insulators} \label{sec:lrsr}
Before moving on to the metallic case, we recap the present understanding of long-range (LR) interatomic forces in insulators. 
To start with, it is useful to recall some basic properties of the force-constants matrix as defined in Eq. (\ref{Eq_FC}). 
First, $\Phi$ is a periodic function of momentum,
\begin{equation}\label{Eq_FC_periodic}
\Phi_{\kappa\alpha,\kappa'\beta}(\textbf{q+G})=\Phi_{\kappa\alpha,\kappa'\beta}(\textbf{q}),
\end{equation}
for any vector ${\bf G}$ that belongs to the reciprocal-space Bravais lattice of the crystal. 
This means that the explicit calculation of $\Phi$ needs to be carried out only within the first Brillouin zone. 
This also means that $\Phi$ can be Fourier-transformed to real space,
\begin{equation}\label{Eq_FC_real_space}
\Phi^l_{\kappa\alpha,\kappa'\beta}=\int_{\rm BZ} [d^3 q] e^{-i{\bf q}\cdot {\bf R}_l} \Phi_{\kappa\alpha,\kappa'\beta}(\textbf{q}).
\end{equation}
The interatomic force constant (IFC) $\Phi^l_{\kappa\alpha,\kappa'\beta}$ describes the force along $\alpha$ on the atom $\kappa$ located in the origin cell that is produced by a displacement along $\beta$ of atom $\kappa'$ in the cell $l$. 
If $\Phi({\bf q})$ were an infinitely differentiable (analytic) function everywhere in 
reciprocal space, the IFCs would be guaranteed to decay exponentially with the interatomic 
distance, $d$.
This is, however, not the case in insulators, where $\Phi({\bf q})$ displays a strongly nonanalytic behavior at the $\Gamma$ ({\bf q}=0) point and its reciprocal-space replicas. 
Such a behavior originates from the diverging nature of the Coulomb kernel, and is responsible for the well-known long-ranged nature of the IFCs. 
At leading order, their asymptotic decay goes like $1/d^3$ because of the dipole-dipole interactions; higher-order multipolar terms that decay as $1/d^4$ (dipole-quadrupole) or faster are generally present as well \cite{Stengel2013,Royo2020}.

For most practical purposes, it is necessary to treat these long-ranged contributions separately from the rest. This is typically done by writing
\begin{equation}\label{Eq_PHi_SR_LR}
\Phi = \Phi^{\rm SR} + \Phi^{\rm LR},
\end{equation}
where the reciprocal-space representation of $\Phi^{\rm SR}$ is required to be analytic and satisfy the periodicity condition, 
Eq.~(\ref{Eq_FC_periodic}). This implies that the real-space IFCs that are associated with $\Phi^{\rm SR}$ via Eq. (\ref{Eq_FC_real_space}) are short-ranged. 
As a consequence, the long-ranged contributions are entirely contained in $\Phi^{\rm LR}$.
It is important to keep in mind that such a requirement leaves a lot of freedom in the choice of how the partition is made. One could even set, for instance, $\Phi^{\rm SR}=0$ and trivially satisfy the aforementioned condition. 
(Such a pathological choice, of course, would bring no practical advantage.)
In most applications of this strategy, $\Phi^{\rm LR}$ is written in reciprocal space as a closed formula, which can be evaluated quickly at every ${\bf q}$ point of the BZ, while $\Phi^{\rm SR}$ is processed via numerical Fourier interpolation.
In the following, we shall discuss some popular strategies that have been developed to achieve an accurate and efficient expression for $\Phi^{\rm LR}$.

\subsubsection{The PCM approach}
The fundamental theory of the interatomic force constants and their long-wavelength behavior was first established by Pick, Cohen and 
Martin \cite{Pick1970} (PCM henceforth). 
The starting point of PCM lies in the observation that the long-ranged character of the IFCs is entirely due to the divergence of the Coulomb kernel at $\Gamma$. 
More specifically, the full Coulomb kernel in reciprocal space reads as
\begin{equation}
 v_{\bf q}(\textbf{G},\textbf{G}') = \delta_{\textbf{G}\textbf{G}'}v_{\bf q}({\bf G}), 
\end{equation} 
where the diagonal elements are given by
\begin{subequations}
\begin{align}
v_{\bf q}({\bf G}) =& v(|\textbf{G+q}|), \\
v(K) =& \frac{4\pi}{K^2}.
\end{align}
\end{subequations}
PCM's strategy consists in separating the Coulomb kernel into the ${\bf G}=0$ term, which diverges
at the $\Gamma$ (${\bf q}=0$) point, and a remainder, $\bar{v}$, that remains 
regular in a neighborhood thereof,
\begin{align}
v_{\bf q}({\bf G}) = \delta_{0\textbf{G}} v(q) + \bar{v}_{\bf q}({\bf G}).
\end{align}
($\bar{v}$ is readily obtained from the full kernel by setting to zero the ${\bf G}=0$ term.)
This naturally leads to a separation of the force-constants near $\Gamma$ into analytic (AN) and nonanalytic (NA) contributions,
\begin{equation}
\Phi = \Phi^{\rm AN} + \Phi^{\rm NA}.
\end{equation}
Here $\Phi^{\rm AN}$ is defined via the same linear-response formula, 
Eq. (\ref{Eq_FC}), as the full $\Phi$, with the only difference that the Coulomb kernel has been replaced with $\bar{v}$. 
$\Phi^{\rm NA}$, on the other hand, enjoys the following compact expression 
\cite{Stengel2013}, 
\begin{equation}
\Phi^{\rm NA}_{\kappa\alpha,\kappa'\alpha'}({\bf q}) =  \bar{\rho}^{*}_{\kappa\alpha}({\bf q}) W({\bf q}) \bar{\rho}_{\kappa'\alpha'}({\bf q}).
\end{equation}
Here $\bar{\rho}_{\kappa\alpha}$ is the macroscopic charge response that is associated with the phonon
mode, Eq. (\ref{Eq_rho_q}), again calculated at the $\bar{v}$ 
level;
$W({\bf q})$ is the head of the screened Coulomb interaction \cite{Royo2021},
and can be expressed as 
\begin{equation}
W({\bf q}) = v(q) \epsilon_{\rm L}^{-1}({\bf q}),
\end{equation}
where $\epsilon_{\rm L}^{-1}({\bf q}) = \epsilon^{-1}({\bf q},{\bf q})$  
is a scalar longitudinal dielectric function, corresponding to the \emph{head} of the inverse dielectric matrix \cite{Macheda2024}.
$\epsilon_{\rm L}^{-1}({\bf q})$, in turn, can be written as
\begin{equation}
\label{eps_l}
\epsilon^{-1}_{\rm L}({\bf q}) = \frac{1}{\epsilon_{\rm L}({\bf q})}, \qquad 
\epsilon_{\rm L}({\bf q}) =  1- v(q) \bar{\chi}({\bf q}),
\end{equation}
where the ``macroscopically unscreened'' susceptibility $\bar{\chi}({\bf q})$ is analytic near $\Gamma$.
This means that $\Phi^{\rm NA}$ can be written in terms of the analytic 
scalar functions $\bar{\chi}$ and $\bar{\rho}$, which can be expanded in powers of ${\bf q}$,
\begin{subequations}
\label{eq:chi_rho_int}
\begin{align}
\bar{\chi} =& -{\bf q} \cdot \bm{\chi}^{\rm mac} \cdot {\bf q} + \cdots, \\
{\Omega} \bar{\rho}_{\kappa \alpha} =& -i {\bf q} \cdot {\bf Z}^*_{\kappa \alpha} - \frac{q_\beta q_\gamma}{2} Q^{(\beta \gamma)}_{\kappa \alpha}+ \cdots.
\end{align}
\end{subequations}
The leading coefficients are, respectively, \cite{Pick1970}  the macroscopic dielectric susceptibility tensor $\bm{\chi}^{\rm mac}$, related to the macroscopic (clamped-ion) dielectric tensor via
$\bm{\epsilon}^{\rm mac} = {\bf I} + 4\pi \bm{\chi}^{\rm mac}$, 
and the BEC tensor ${\bf Z}^*_{\kappa \alpha}$.
For completeness, in Eq.~\eqref{eq:chi_rho_int} we also include the dynamical quadrupole tensor ${\bf Q}_{\kappa \alpha}$, as it has received considerable attention recently~\cite{Vogl1976,Stengel2013,Verdi2015,Sjakste2015,Brunin2020,Park2020}.
At small momenta one then obtains the well-known limiting behavior, 
\begin{equation}\label{Eq_Phi_NA_q0}
\begin{split}
\Phi_{\kappa\alpha,\kappa'\beta}^{\rm NA}({\bf q}\rightarrow 0) &=
4\pi\Omega 
\frac{\bar{\rho}^{\mathbf{q} \, *}_{\kappa\alpha} \,
\bar{\rho}^\mathbf{q}_{\kappa'\beta}}{ q^2 \epsilon_{\rm L}(\mathbf{q})}\bigg|_{\mathbf{q}\rightarrow 0}\\
&=
\frac{4\pi}{\Omega} \frac{(\hat{q}\cdot\mathbf{Z}_\kappa^*)_\alpha 
(\hat{q}\cdot\mathbf{Z}_{\kappa'}^*)_\beta}
{\hat{{q}}\cdot\boldsymbol{\epsilon}^{\rm mac}\cdot \hat{{q}}},
\end{split}
\end{equation}
where we have defined $\hat{q} = {\bf q}/q$.

The main drawback of the PCM approach lies in that the AN and NA terms do not lend themselves to a direct interpretation as short-range and long-range contributions to the IFC. Indeed, the $\Phi^{\rm AN}$ constructed this way breaks the basic requirement of periodicity in reciprocal space, Eq. (\ref{Eq_FC_periodic}): 
as it stands it cannot be Fourier transformed to real space. Moreover, $\Phi^{\rm AN}$ is analytic only in a vicinity of ${\bf q}=0$, and is still nonanalytic at any other primitive reciprocal-space vector.
For the same reason, the function $\epsilon_{\rm L}({\bf q})$ defined in 
Eq. (\ref{eps_l}) cannot be directly interpreted as a macroscopic dielectric function, and cannot be incorporated in the Maxwell equations as it stands.
This means that, within PCM, the construction of $\Phi^{\rm LR}$ involves two separate steps. First, via the above derivations one determines the exact nonanalytic behavior of $\Phi$ near $\Gamma$. Second, one uses this information to build a model for the long-range interatomic forces, whose reciprocal-space expression must reproduce the nonanalytic behavior of  $\Phi^{\rm NA}$
near $\Gamma$, and be periodic. 

While this two-step approach has been very successful for decades, the interest for strategies that provide the range-separated IFC's (and, as a byproduct, the relevant macroscopic physical properties of the crystal) at once 
has been steadily growing.
Such a necessity is primarily fueled by the ongoing popularity of low-dimensional systems, and by the benefit (in terms of computational accuracy) of generalizing the dipole-dipole formula to higher multipolar orders. In all these cases, the PCM strategy becomes cumbersome to implement properly.
The recently developed range-separation method presents an effective way beyond this limitation; we shall review it in the following paragraphs.
\subsubsection{The range-separation method}
The separation between SR and LR is performed by splitting the bare Coulomb kernel into two parts,
\begin{equation}
v = v^{\rm SR} + v^{\rm LR}.
\end{equation}
Both parts are, as the original kernel, diagonal operators that act on the full plane-wave basis set, e.g.,
\begin{equation}
 v^{\text{LR}}_{\bf q}(\textbf{G},\textbf{G}') = \delta_{\textbf{G}\textbf{G}'}v^{\text{LR}}_{\bf q}({\bf G}).
\end{equation} 
Crucially, at difference with PCM, here we impose the additional periodicity condition, Eq. (\ref{Eq_FC_periodic}), on both $v^{\rm SR}$ and $v^{\rm LR}$. This way, we ensure that $v^{\rm SR}$ is an analytic function of ${\bf q}$ over the entire reciprocal space, and therefore describes a truly short-ranged interaction once Fourier-transformed to real space.
A sufficient condition to guarantee periodicity is to use, just like in the case  of the full kernel,
\begin{equation}
v^{\text{SR,LR}}_{\bf q}({\bf G}) = v^{\text{SR,LR}}(|\textbf{G+q}|).
\end{equation}
Then, in close analogy to the nanosmoothing \cite{Junquera2007} technique, the range separation can be operated by writing, 
\begin{subequations}
    \begin{align}
        v^{\text{LR}}(K) =& f(K) v(K), \\
        v^{\text{SR}}(K) =& [1-f(K)] v(K),
    \end{align}
\end{subequations}
where $f(K)$ is an appropriate Fourier-filtering function.
The specific choice of $f(K)$ leaves room for some freedom. 
In Ref.~\citenum{Royo2021}, a Gaussian function was proposed, 
\begin{equation}\label{Eq_f_gaussian}
    f(K) = \exp\left(-\frac{K^2L^2}{4}\right),
\end{equation}
where $L$ is a range-separation parameter that is used to define what one regards as ``macroscopic physics''  in the present context \cite{Royo2021}.
One can easily verify that $v^{\rm SR}$ is then a regular function of ${\bf q}$, as it should.
The procedure that one uses at this point to separate the SR and LR parts of the force-constants matrix 
follows similar guidelines as in the PCM method.
The main difference is that here we can directly write the final result, 
Eq. (\ref{Eq_PHi_SR_LR}), in one shot.
\PhiSR{} is defined and calculated with the $v^{\text{SR}}$ Coulomb kernel, where the diverging macroscopic part (responsible for the long-range interactions) has been suppressed. 
The LR term, in turn, is written 
\cite{Stengel2013} in the following form, 
\begin{equation}
\label{eq:philr}
\Phi^{\rm LR}_{\kappa\alpha\kappa'\alpha'}({\bf q}) =  \rho^{\rm SR*}_{\kappa\alpha}({\bf q}) W^{\text{LR}}({\bf q}) \rho^{\rm SR}_{\kappa'\alpha'}({\bf q}).
\end{equation}
Here $\rho$ is the macroscopic charge response that is associated with the phonon
mode,
$W^{\text{LR}}({\bf q})$ is the macroscopic screened Coulomb interaction \cite{Royo2021},
and can be expressed as 
\begin{equation}
\label{eq:wlr}
W^{\text{LR}}({\bf q}) = v^{\text{LR}}(q) \epsilon_{\text{LR}}^{-1}({\bf q}).
\end{equation}
where 
\begin{equation}
\label{eq:elr}
\epsilon_{\text{LR}}({\bf q}) = 1 - v^{\text{LR}}(q) \chi^{\rm SR}({\bf q})
\end{equation}
is the macroscopic dielectric function; $\chi^{\rm SR}({\bf q})$, in turn, is the macroscopic  susceptibility calculated at the SR level.

In full generality, both $\rho^\SR$ and $\chi^{\rm SR}$ are analytic in the long-wavelength limit, where they enjoy the following Taylor expansion,
\begin{subequations}
\label{eq:chi_rho_int2}
\begin{align}
{\chi}^{\rm SR} =& -{\bf q} \cdot \bm{\chi}^{\rm mac} \cdot {\bf q} + \cdots, \\
{\Omega} {\rho}^{\rm SR}_{\kappa \alpha} =& -i {\bf q} \cdot {\bf Z}^*_{\kappa \alpha} - \frac{q_\beta q_\gamma}{2} Q^{(\beta \gamma)}_{\kappa \alpha}+ \cdots.
\end{align}
\end{subequations}
Note that the leading terms in the expansion are the same in 
Eq. (\ref{eq:chi_rho_int2}) and in their PCM counterparts. The proof that the dynamical quadrupoles are insensitive to the range-separation parameter was first provided by Martin in his seminal work on piezoelectricity \cite{Martin1972}. In the context of the charge response to a phonon, the $L$-dependence starts to kick in at the octupolar level, although the clamped-ion flexoelectric coefficient, given by the basis sum of the octupoles, \cite{Resta2010,Hong2011} remains unaffected. 
(In Ref. \cite{Martin1972} the range separation was performed via a Yukawa-like potential, which leads to an $f(K)$ in a  Lorentzian form instead of the Gaussian filter of
Eq. (\ref{Eq_f_gaussian}); the two choices lead to the same conclusions.)
We stress that the range-separation approach is essential to obtaining a real-space representation of the charge-density response to an atomic displacement 
\cite{Stengel2013} and for defining, in turn, its multipolar moments.

By comparing the PCM and range-separation approaches one quickly realizes that most of the formulas look exactly the same, except for the superscript labels on some of the symbols. The main difference between the two consists in how the full Coulomb kernel is split, which in the range-separation case preserves the periodicity in reciprocal space; the advantages of doing so have already been discussed at length in the above paragraphs. 
At first sight, the involvement of an arbitrary length parameter, $L$, in the derivations may appear as a limitation of the latter method. Such a parameter is, however, ubiquitous in problems involving electrostatics (e.g., the Ewald method). 
Indeed, the choice of a length scale is implicit in all circumstances where macroscopic physical information needs to be extracted from microscopic potentials or charge densities. The conceptual foundation of the time-honored  macroscopic Maxwell equations rests precisely on local averaging over some volume that is large compared to the lattice spacing. The ``nanosmoothing'' \cite{Junquera2007} or ``macroscopic averaging'' \cite{Baldereschi1988} techniques, closely related to the range-separation approach described here, can be regarded as an implementation of these ideas in a modern context.
One should keep in mind that $L$ has no impact on the total IFCs: \PhiLR{} and \PhiSR{} bear an equal and opposite $L$-dependence that cancels out exactly once they are summed over.

\subsection{Metallic crystals}\label{Sec_theory_metals}
While the above derivation has been historically established for
insulators, it can be carried out just as well in metallic crystals.
There is an additional complication here, related to the abrupt change in the orbital population at the Fermi level. Such a sharp feature may induce additional nonanalytic terms in the force constants (e.g., Kohn anomalies) that are unrelated to electrostatics. 
To circumvent this issue, we assume a finite electronic temperature henceforth, 
and thereby eliminate the anomalies in the static response. 
Within such an assumption, the response functions of 
Eq.~(\ref{eq:chi_rho_int}) are characterized by the following limiting behavior, 
\begin{subequations}
\label{eq:chi_rho_met}
\begin{align}
{\chi}^{\rm SR} =& -\frac{C_{\rm Q}}{\Omega} + \mathcal{O}(q^2),\label{eq:chi_rho_met_a} \\
\Omega {\rho}^{\rm SR}_{\kappa \alpha} =& M_{\kappa \alpha} -i {\bf q} \cdot {\bf Z}^{\rm stat}_{\kappa \alpha} + \cdots. \label{eq:chi_rho_met_b}
\end{align}
\end{subequations}
The main difference with the insulating case is that both functions generally tend to a finite constant in the long-wavelength limit. 
These monopolar terms might be surprising at first sight, as they seemingly violate charge neutrality. However, recall that the ``SR'' quantities are defined within the modified Coulomb kernel where macroscopic electrostatic effects have been suppressed.
Such long-range fields are precisely the physical ingredients that prevent accumulation of a net charge over extended region of the crystal, and enforce exact particle conservation in the ${\bf q}\rightarrow 0$ limit.

As we shall see shortly, the parameter $C_{\rm Q}$ has the physical interpretation of a \emph{quantum capacitance}. 
The parameters $M_{\kappa \alpha}$, on the other hand, embody the monopolar charge response to a lattice-periodic (zone-center) distortion of the crystal. 
In many crystals of sufficiently high symmetry, $M_{\kappa \alpha}$ vanish identically. Then, the parameters ${\bf Z}^{\rm stat}_{\kappa \alpha}$ describe the charge response to a phonon at leading order; they can be regarded \cite{Macheda2022} as the static counterpart of the dynamical Born charges \cite{Saitta2008,Bistoni2019,Binci2020,Dreyer2022} in metals. 

It is important to note that, by incorporating Eq.~(\ref{eq:chi_rho_met}) into Eq.~(\ref{eq:philr}), \PhiLR{} becomes analytic at ${\bf q}=0$. 
This result constitutes the main qualitative difference between insulators and metals in our context: in the latter, the free carriers screen the long-range electrostatic interactions completely within the static regime.
This means that, at the formal level, the explicit separate treatment of \PhiLR{} is unnecessary.
Yet, as mentioned above, recent work \cite{Macheda2022} has demonstrated that the numerical interpolation of the IFCs and electron-phonon matrix elements might still require some care. Indeed, in typical doped semiconductors, free-carrier screening is effective only in a narrow region of reciprocal space near $\Gamma$. Such a behavior, unless accounted for explicitly, requires a dense mesh in momentum space to be resolved properly,  severely hampering computational efficiency.
For this reason, it is important to clarify the physics of the various parameters entering Eq.~(\ref{eq:chi_rho_met}), which we shall do in the following sections.

\subsection{Quantum capacitance}
Since the main qualitative difference between insulators and metals resides in the long-wavelength behavior of the charge susceptibility,  
we shall frame our discussion in this section around the \emph{quantum capacitance} $C_{\rm Q}$, which describes the limiting value of $\chi^{\text{SR}}$ at the zone center.
In particular, from Eqs.~(\ref{eq:chi_q}) and (\ref{Eq_capacitance}) we obtain
\begin{equation}
\label{eq:C_DFPT}
C_{\rm Q} = -\Omega \chi^{\rm SR}({\bf q}=0) = \int_\text{BZ} [d^3 k]  \sum_{m} f'_{m\bf k} \langle u_{m \bf k}| \hat{\mathcal{H}}^{\varphi} |u_{m \bf k} \rangle,
\end{equation}
where $f'_{m\bf k}=\partial f_{m\bf k}/\partial \epsilon_{m\bf k}$ is the energy derivative of the Fermi distribution function. 
After observing that the external scalar potential perturbation at the zone center reduces to $H^\varphi=-1$, $C_{\rm Q}$ can be equivalently written as  
\begin{equation}
\label{eq:C}
C_{\rm Q} = \frac{\partial N}{\partial \mu},
\end{equation}
i.e., the derivative of the total number of electrons in the unit cell $N$ in response to a uniform shift of the Fermi level $\mu$.
(The derivative is intended to be taken with the full Coulomb kernel replaced by $v^{\rm SR}$, which regularizes 
the otherwise diverging electrostatic energy of the charged crystal).
Note that in most first-principles codes it is easier to control $N$ rather than $\mu$, which might favor the converse formula,
\begin{equation}
\label{eq:C_inv}
C_{\rm Q}^{-1} = \frac{\partial \mu}{\partial N}.
\end{equation}

When SCF fields are neglected, $C_{\rm Q}$ reduces to the density of states at the Fermi level broadened by the Fermi factor,
\begin{equation}
\label{eq:C_lind}
    C_0 = -\int_\text{BZ} [d^3 k]  \sum_{m} f'_{m\bf k}.
\end{equation}
$C_0$ relates to the ${\bf q}\rightarrow 0$ limit of the static Lindhard susceptibility via $C_0=-\Omega \chi^{\rm Lind}({\bf q}=0)$; thus, we shall refer to $C_0$ as ``Lindhard capacitance'' henceforth.
Since the Fermi occupation function is monotonically decreasing, $C_0$ is positive by construction.
By switching on the SCF fields, there is an additional contribution from Hartree local fields and XC effects, which adds to $C_0$ via a series capacitor formula,
\begin{equation}
\label{eq:cq}
C_{\rm Q}^{-1} = C_0^{-1} + C_{\rm Hxc}^{-1}.
\end{equation}
Such quantum corrections, embodied by $C_{\rm Hxc}$, have received considerable
attention 
several years ago in the context of
confined electron gases \cite{Ashoori1992,Buttiker1993,Eisenstein1994,Kopp2009,Li2011,Junquera2019}. 
Interestingly, they were found to be of crucial importance
in the dilute limit, where they may induce a transition to 
a \emph{negative electron compressibility} regime. \cite{Kopp2009,Junquera2019} 
(In the context of Refs.~\citenum{Kopp2009,Junquera2019}, the Lindhard capacitance plays the role of the kinetic-energy contribution.)

Based on the above considerations, the quantum capacitance appears to be an important physical property of the metallic crystal.
Nevertheless, 
recall that all quantities calculated within the SR kernel must be treated with care, and their possible dependence on the inherent arbitrariness in the definition of $v^{\rm SR}$ verified in each case. We show in Appendix~\ref{app:int_pol} that the following result holds in full generality,
\begin{equation}
\label{eq:C_Hxc}
C_{\rm Hxc}^{-1} = \bar{C}_{\rm Hxc}^{-1} + C_{\rm geom}^{-1}, 
\end{equation}
where $\bar{C}_{\rm Hxc}$ is the value calculated within the
PCM convention (which corresponds to neutralizing the excess charge with a uniform compensating background), and $C_{\rm geom}$ refers to an additional ``geometric'' contribution that arises when other conventions are adopted. 
In particular, if we use the range-separation approach of 
Eq.~(\ref{Eq_f_gaussian}), we obtain $C_{\rm geom}^{-1} = \pi L^2 / \Omega$, where $L$ is the range separation parameter.

The $L$-dependence of the static capacitance might be surprising at first sight. 
To see why this happens, recall that the (inverse) quantum capacitance is the second derivative of the energy with respect to the net charge in the cell. 
As we said, the electrostatic energy of an extended charged system diverges, so one needs to regularize the result somehow. Once the divergence is removed, however, the left-over constant depends on the details of the regularization procedure. 
Within the formalism of Ref.~\citenum{Royo2019}, one multiplies the Coulomb kernel in real space by an error function to make it short-ranged (and hence a regular function of ${\bf q}$ once transformed back to reciprocal space). In other words, one truncates the Coulomb tail beyond some radius $L$: This means that the larger $L$ is, the larger the portion of charged crystal will be accounted in the calculation of the SR energy. (And indeed, $L$ defines the length scale that separates what one regards as macroscopic potentials from local fields~\cite{Royo2021}.) Eventually the capacitance vanishes in the limit $L\rightarrow \infty$, when the SR kernel coincides with the full kernel and charge neutrality is enforced exactly.

\subsection{Sum rules for the charge response\label{sec:sum_rule}}

The capacitance discussed in the previous section is intimately related to the physics of the charge response to a phonon as well. To see that, it is useful to observe that the Fermi level and the particle number are conjugate variables (more specifically, Legendre pairs). 
This means that, by switching from the full Coulomb kernel to the SR version, we have effectively changed the thermodynamic ensemble of the electronic subsystem from canonical (fixed electronic temperature and particle number) to grand-canonical (fixed chemical potential).
Based on such an analogy one can write, for example,
\begin{equation}
\label{eq:mka}
M_{\kappa \alpha} = \frac{\partial N}{\partial \tau_{\kappa \alpha}}\Big|_{\mu} = -\frac{\partial \mu}{\partial \tau_{\kappa \alpha}}\Big|_{N} C_{\rm Q}.
\end{equation}
That is, the total charge induced by an atomic displacement at fixed $\mu$ is equal to the Fermi-level shift at fixed $N$ times the capacitance $C_Q$.  
Since there is no Fermi level shift associated with a rigid translation of the crystal, we can immediately deduce the following sum rule for the monopolar terms,
\begin{equation}
\label{eq:sum_rule1}
\sum_\kappa M_{\kappa \alpha} = 0.
\end{equation}

To derive a sum rule for the ``static charges'' appearing at first order in ${\bf q}$, recall the following result~\cite{Schiaffino2019} for the sublattice sum of the charge-density response to a phonon, $\rho_{\kappa\alpha}^{\bf q}({\bf r})$, 
\begin{equation}
\label{eq:metric_rho}
\sum_\kappa \rho_{\kappa\alpha}^{\bf q}({\bf r}) = \rho^{(\alpha)}_{\bf q}({\bf r}) + 
 \Delta \rho^{\alpha}_{\bf q}({\bf r}),
\end{equation} 
where $\rho^{(\alpha)}_{\bf q}({\bf r})$ is the first-order density response to a \emph{metric wave} perturbation \cite{Stengel2018,Schiaffino2019}, and $\Delta \rho^{\alpha}_{\bf q}({\bf r})$ is a ``geometric'' contribution, i.e., defined in terms of the ground-state wavefunctions only. 
(We drop the ``SR'' superscript on the response functions and assume that all quantities are defined and calculated within SR electrostatics in the remainder of this section.) 

It can be shown (see Appendix~\ref{app:delta_n}) that the geometric contribution has the same form as the insulating case \cite{Schiaffino2019}
\begin{equation}
\begin{split}
\label{eq:delta_n}
\Delta \rho^{(\alpha)}_{\bf q}({\bf r})=  -\frac{ \partial \rho({\bf r}) }{\partial r_\alpha} - i q_\alpha \rho({\bf r}),
\end{split}
\end{equation}
where $\rho({\bf r})$ is the ground-state density. 
On the other hand, the metric-wave response is related to the uniform strain~\cite{Hamann2005} response via~\cite{Schiaffino2019} 
\begin{equation}
\label{eq:hamann}
 \rho^{(\alpha)}_{\bf q}({\bf r}) = i q_\beta \rho^{\eta_{\alpha\beta}}({\bf r}).
\end{equation}
The integral of Eq.~(\ref{eq:metric_rho}) over a primitive cell vanishes, since the crystal is charge neutral in the ground state. 
Thus, by integrating both sides of Eq.~(\ref{eq:metric_rho}) over a primitive cell, and by using Eq.~(\ref{eq:hamann}), Eq.~(\ref{eq:chi_rho_met}) and Eq.~(\ref{eq:sum_rule1}), we obtain
\begin{equation}
-iq_\beta \sum_\kappa Z_{\kappa \alpha}^{\rm stat,\beta} + \mathcal{O}(q^2) = i q_\beta M^{\eta_{\alpha\beta}} + \mathcal{O}(q^2),
\end{equation}
where we have indicated as $M^{\eta_{\alpha\beta}}$ the net charge accumulated in the cell by a uniform strain $\eta_{\alpha\beta}$.
By focusing our attention to the leading order in the momentum, and by accounting for the negative sign of the electron charge we arrive at
\begin{equation}
\label{eq:sum_rule2}
\sum_\kappa Z_{\kappa \alpha}^{\rm stat,\beta} =
\frac{\partial N}{\partial \eta_{\alpha\beta}}\Big|_{\mu} =-\frac{\partial \mu}{\partial \eta_{\alpha\beta}}\Big|_{N} C_{\rm Q},
\end{equation}
where $\eta_{\alpha\beta}$ indicates the derivative with respect to a uniform strain perturbation~\cite{Hamann2005}.
Note the essentially perfect analogy between  Eqs.~(\ref{eq:sum_rule2}) and (\ref{eq:mka}): the only difference on the right-hand side consists in the replacement of a phonon with a uniform strain derivative.

The sum rule Eq.~(\ref{eq:sum_rule2}) belongs to a broad class of relations that link the physics of macroscopic deformations to the long-wavelength behavior of lattice distortions.
Notable examples are the microscopic theory of the elastic tensor, already established by Born and Huang in the 1950s \cite{Born1954} , and the pioneering theory of piezoelectricity of Martin \cite{Martin1972}.
Both theories were revisited in recent times within a modern first-principles framework, and generalized to the treatment of flexoelectricity \cite{Hong2013,Stengel2013}.
Most of these works have primarily focused their attention to insulators, while the sum rules of  Eqs.~(\ref{eq:mka}) and (\ref{eq:sum_rule2}) are specific to metals and a key result of this work.

\subsection{Screened charges and potentials \label{sec:screened}}

Given the central role played by the quantum capacitance, $C_{\rm Q}$, in the charge response to a phonon, $\rho^{\rm SR}_{\kappa \alpha}({\bf q})$, it can be expected that the dependence of the former on the range separation parameter $L$ propagates to the latter as well.
To see that this is indeed the case, we write \cite{Royo2021} 
\begin{equation}
\label{rhoscr}
\begin{split}
\rho^{\rm SR}_{\kappa \alpha} = & \epsilon_{\rm LR} \, \rho_{\kappa \alpha} = W_{\rm LR}^{-1} \, \varphi_{\kappa \alpha},
\end{split}
\end{equation}
where $\epsilon_{\rm LR}$ and $W_{\rm LR}$ are the macroscopic dielectric function and the screened Coulomb interaction defined in Section~\ref{sec:lrsr}, while $\rho_{\kappa \alpha}$ and $\varphi_{\kappa \alpha}$ are, respectively, the (fully) screened charge and electrostatic potential response to the phonon.
All quantities appearing in Eq.~(\ref{rhoscr}) are scalar functions of the momentum ${\bf q}$, and coincide with the ``small-space'' representation of macroscopic electrostatics as detailed in Ref.~\cite{Royo2021}. [The present Eq.~(\ref{rhoscr}) corresponds to Eq.~(48a) therein.]
As a consequence of Eq.~(\ref{eq:wlr}), the following relation trivially holds, 
\begin{equation}
\label{v_rho_scr}
    \varphi_{\kappa \alpha} = v_{\rm LR} \, \rho_{\kappa \alpha}.
\end{equation}

The convenience of working with the screened functions $\varphi_{\kappa \alpha}$ and $\rho_{\kappa \alpha}$  lies in the fact that $\rho_{\kappa \alpha}$ is unaffected by the arbitrariness of $v_{\rm LR}$, and due to Eq.~(\ref{v_rho_scr}), 
$\varphi_{\kappa \alpha}$ is also unaffected at the lowest orders in ${\bf q}$ that are relevant to the present discussion. 
Thus, we can write
\begin{align}
    \label{eq:V_exp}
    \varphi_{\kappa \alpha}({\bf q}) =&  \varphi_{\kappa \alpha}^{(0)} - iq_\beta \varphi_{\kappa \alpha}^{(1,\beta)} + \mathcal{O}(q^2), \\
     \label{eq:rho_exp}
    \rho_{\kappa \alpha}({\bf q}) = & q^2 \frac{\varphi_{\kappa \alpha}^{(0)}}{4\pi}  - iq^2 q_\beta
     \frac{\varphi_{\kappa \alpha}^{(1,\beta)}}{4\pi} + \mathcal{O}(q^4),
\end{align}
where neither $\varphi^{(0)}$ or $\varphi^{(1)}$ depend on $L$.
We can relate Eq.~(\ref{eq:rho_exp}) to the standard formulas that are used in the multipolar expansion of $\rho_{\kappa \alpha}({\bf q})$ \cite{Martin1972,Stengel2013} 
involving, e.g., the dynamical quadrupole $Q^{(\beta\gamma)}_{\kappa \alpha}$ and octupole $O^{(\beta\gamma \lambda)}_{\kappa \alpha}$ tensors; specifically, $Q_{\kappa \alpha} = - 2\Omega {\varphi_{\kappa \alpha}^{(0)}}/(4\pi)$, where ${Q_{\kappa \alpha} \equiv Q^{(\beta\gamma)}_{\kappa \alpha}\delta_{\beta\gamma}}$ and $O^\beta_{\kappa \alpha} = -6\Omega \varphi_{\kappa \alpha}^{(1,\beta)}/(4\pi)$, where $O^\beta_{\kappa \alpha}\equiv O^{(\beta\beta \beta)}_{\kappa \alpha}$ [or equivalently $O^\beta_{\kappa \alpha}\equiv 3 O^{(\beta\gamma \gamma)}_{\kappa \alpha}$] \cite{Stengel2013}. 
Clearly, the screened charge-density response kicks in at the quadrupolar order, and meets the strict requirement that all electrostatic multipoles of an isolated atomic displacement must vanish.
This means that all components of the Cartesian multipolar tensors vanish except for their traces, which cannot generate long-range fields.
(A multipolar expansion of the charge on a spherical harmonic basis yields a null result at any order.)

At this point, we are only left with the task of determining the macroscopic dielectric function, which is readily given by Eq.~(\ref{eq:elr}) as
\begin{equation}
\label{eq:eps_LR}
    \epsilon_{\rm LR} = \frac{4\pi}{q^2} \frac{C_{\rm Q}}{\Omega} + \mathcal{O}(q^0).
\end{equation}
This allows us to write down the leading monopolar and dipolar terms in the long-wavelength expansion of $\rho^{\rm SR}_{\kappa\alpha}$ in terms of the screened potentials introduced earlier in this section. In particular, by combining Eq.~(\ref{eq:chi_rho_met}) with Eqs. (\ref{rhoscr})-(\ref{eq:eps_LR}) we obtain
\begin{subequations}
\label{eq:mz_v}
\begin{align}
    M_{\kappa \alpha} =& \varphi_{\kappa \alpha}^{(0)} \, C_{\rm Q}, \\
    Z^{\rm stat,\beta}_{\kappa \alpha} =& \varphi^{(1,\beta)}_{\kappa \alpha} \, C_{\rm Q}.
\end{align}
\end{subequations}
This result is nicely consistent with Eq.~(\ref{eq:mka}), after observing that the relative shift of the Fermi level (with respect to the electrostatic reference) at fixed particle number must be equal to (minus) the shift of the electrostatic potential at fixed Fermi level. 
It also allows us to rewrite our second sum rule, Eq.~(\ref{eq:sum_rule2}), in a form that only involves screened electrostatic potentials, and is therefore independent of the quantum capacitance,
\begin{equation}
\label{eq:sumrule_alt}
    \sum_\kappa \varphi^{(1,\beta)}_{\kappa \alpha} = -\frac{\partial \varphi}{\partial \eta_{\alpha\beta}}\Big|_{N}.
\end{equation}
Most importantly, Eq.~(\ref{eq:mz_v}) allows us to answer the question we set to ourselves at the beginning of this section: both $M_{\kappa\alpha}$ and \Zstat{} are indeed affected by the same ambiguity as the quantum capacitance, originating from the arbitrary partition of the kernel between macroscopic and microscopic components.
Note that both sum rules of Section~\ref{sec:sum_rule} are immune from this issue: this is manifest from Eq.~(\ref{eq:sumrule_alt}).

The above results imply that, if the range-separation approach is used, both $M_{\kappa\alpha}$ and \Zstat{} vanish at large-$L$, consistent with the fact that $v_{\rm SR}$ coincides with the full kernel in that limit.
In practice, this means that \PhiLR{} can be made arbitrarily small by transferring an arbitrary portion of the ``long-range'' force constants into \PhiSR{}.
This might, again, appear surprising; however, recall that \PhiLR{} in a finite-temperature metal is an analytic function of ${\bf q}$. Thus, at the formal level there is no reason why it should be treated separately from \PhiSR{}, hence the freedom to transfer pieces back and forth between \PhiSR{} and \PhiLR{}.
Choosing the form of \PhiLR{} becomes then a matter of computational convenience, as it may allow (for some specific choices of $L$ and of the reference potential) to accelerate the calculation of the electron-phonon interactions in a vicinity of the zone center.

Before closing, it is important to emphasize that the screened susceptibility $\chi$, just like $\rho_{\kappa\alpha}$, is unaffected by the arbitrariness in the definition of $C_{\rm Q}$. 
In other words, in the relation [see Ref.~\cite{Royo2021}, Eq.(48b)],
\begin{equation}
\label{eq:chi_SR_chi}
    \chi^{\rm SR} =  \epsilon_{\rm LR} \, \chi.
\end{equation}
the respective ambiguities in $\chi^{\rm SR}$ and $\epsilon_{\rm LR}$ cancel out exactly at any order.
More explicitly, using Eqs.~(\ref{eq:chi_rho_met_a}) and (\ref{eq:elr}), one can write $\chi$ in terms of the quantum capacitance as
\begin{equation}
\label{eq:screen_chi}
    \begin{split}
      {\chi}({\bf q}) =& -v^{-1}_{\rm LR} + \Omega {C}_{\rm Q}^{-1} v^{-2}_{\rm LR} + \mathcal{O}(q^6) \\
      =& -\frac{q^2}{4\pi} + \Omega \bar{C}_{\rm Q}^{-1} \left( \frac{q^2}{4\pi}\right)^2 +  \mathcal{O}(q^6).
    \end{split}
\end{equation}
Note that $\chi$ must have only even powers of \textbf{q} and the zeroth-order term vanishes.
The leading-order term in the charge susceptibility is material independent: its analytic form reflects the perfect screening of any external charge perturbation in the long-wavelength limit. 
It is easy to verify that the two expressions for $\chi$ provided in Eq.~(\ref{eq:screen_chi}) also coincide at $\mathcal{O}(q^4)$, once we observe that $v_{\rm LR} = q^2/4\pi (1 + q^2L^2/4) + \mathcal{O}(q^6)$.

\subsection{Reference potential dependence \label{sec:ref_dep}}

In addition to the issues discussed in the previous paragraphs, our sum rule for the \Zstat{} reveals another source of concern with respect to their physical significance, which we explain in the following. 
The charge response to a strain, appearing on the right-hand side of Eq.~(\ref{eq:sum_rule2}), can be equivalently written as
\begin{equation}
\label{eq:defpot}
    \frac{\partial N}{\partial \eta_{\alpha\beta}}\Big|_{\mu} = \int [d^3 k] \sum_m
     f'_{m\bf k} \frac{\partial \epsilon_{m\bf k}}{\partial \eta_{\alpha\beta}}\Big|_{\mu},
\end{equation}
i.e., in terms of the energy derivatives of the Fermi-Dirac occupation function and the strain derivatives of the band energies.
The latter are known in the literature as \emph{relative} deformation potentials (RDP) \cite{Vandewalle1989,Resta1990,Stengel2015}, since the eigenvalues are not meaningful in an \emph{absolute} sense, but rather referred to the energy zero that is set internally by the code.
Choosing the energy zero in a periodic crystal is, in practice, largely a matter of convention, and different electronic-structure packages adopt different prescriptions. For this reason, any physical property that explicitly depend on such choice needs to be treated with caution.
This difficulty is commonly known in the literature as \emph{reference potential ambiguity} \cite{Royo2022}.

To illustrate how the reference potential ambiguity affects both \Zstat{} and the sum rule, 
Eq.~(\ref{eq:sum_rule2}), in the following we shall make an explicit comparison between two popular choices.
On one hand, we consider the default convention in the {\sc abinit} code \cite{Abinit_1} (referred to here as \Vloc{}), which consists in setting to zero the cell average of $V^{\rm H} + V^{\rm loc}$, i.e., the sum of the  Hartree (H) potential and the local (loc) part of the atomic pseudopotential \cite{Bruneval2014}).
On the other hand, we consider the prescription of Refs. 
\cite{Stengel2013,Stengel2014,Royo2019,Royo2022},
referred to as \Velec{} henceforth, which consists in setting to zero the cell average of the electrostatic potential. The latter is defined as $V^{\rm H} + V^{\rm pc}$, where ``pc'' refer to a set of classical point charges located at the lattice sites and equal to the total number of electrons of each pseudoatom.

It is easy to show (and we shall prove this point numerically 
in Sec.~\ref{sec:res}) that the values of \Zstat{} are indeed affected by a change of reference (e.g., by switching from the \Vloc{} to the \Velec{} convention). 
To see this, note that the change of reference boils down to superimposing a set of spherically symmetric and neutral charge distributions (corresponding, in this case, to the Laplacian of the local pseudopotential minus the ionic point charge) to the atomic Hamiltonian. 
These additional rigid charges are centered at the atomic sites, and therefore contribute to the screened potential response to a phonon as \cite{Royo2021}
\begin{equation}
V^{\rm loc}_{\kappa \alpha}({\bf q}) - V^{\rm pc}_{\kappa \alpha}({\bf q}) = -iq_\alpha \frac{F''_\kappa}{2 \Omega},
\end{equation}
where $F''_\kappa$ is related to the quadrupolar moment of $\rho^{\rm loc}$. 
This implies that, if we use \Vloc{} instead of \Velec{} as our definition of the screened potential, we obtain a correction at first order in ${\bf q}$,
\begin{equation}
    \Delta \varphi^{(1,\beta)}_{\kappa \alpha} = \delta_{\alpha \beta} \frac{F''_\kappa}{2 \Omega}.
\end{equation}
This difference propagates to the ``macroscopically unscreened'' response properties by applying the ${\bf G}\neq 0$ prescription to one or the other reference, and results in a discrepancy of 
\begin{equation}
\label{eq:ref_dep}
  \Delta Z^{\rm stat,\beta}_{\kappa \alpha} =   \Delta \varphi^{(1,\beta)}_{\kappa \alpha} \, C_{\rm Q}.
\end{equation}
Note that the sum rule using either convention is exact, provided that the same reference is used in the calculation of the \Zstat{} and in the strain response; indeed, the impact of the change of reference on the uniform strain Hamiltonian corresponds precisely to the lattice sum of $\Delta \varphi$ [see Ref.~\citenum{Royo2021}, Eq. (C5) and (C8)].

Before closing this section, it is useful to mention that the use of pseudopotentials in itself can be problematic, even if we were to ``trust'' the electrostatic potential (e.g., in some application where it might be relevant). Nonlocal pseudopotentials are not meant to reproduce the quadrupolar moment of the spherical all-electron atom, and corrections are usually needed in order to reproduce the correct form factor of the ionic charge.
This issue has already been pointed out in the case of the flexoelectric tensor \cite{Hong2011};  in principle, a similar correction would also be necessary for reproducing the all-electron value of \Zstat{} in a pseudopotential calculation, 
\begin{align}
\label{eq:delta_z}
\Delta Z^{\rm stat}_{\kappa\alpha} =& -\frac{4\pi C_{\rm Q}}{6\Omega}  \Delta O^\beta_{\kappa\alpha}, \\
\Delta O^\beta_{\kappa\alpha}=& 4\pi \delta_{\alpha\beta} 
    \int dr\, r^4 \left[\rho_\kappa^{\text{AE}}(r)-\rho_\kappa^{\text{PS}}(r)\right],
    \label{eq_delta_z_2}
\end{align}
where $n_\kappa^{\text{AE}}(r)$ and $n_\kappa^{\text{PS}}(r)$ are the all-electron and pseduopotential charge densities.
Since the electronic core charge is negative, Eq.~(\ref{eq:delta_z}) always leads to a positive contribution to \Zstat{}, which can become important in heavier elements.
The fact that the quadrupolar form factor of the core affects the value of \Zstat{} further questions the physical significance of the latter in electron-phonon and lattice-dynamical calculations, where the core electrons are usually regarded as chemically inert.

\section{Computational approach \label{sec:comp}}
In this section, we discuss the code implementation 
of the quantities introduced in Section \ref{sec:form}. 
In particular, we will focus on the uniform strain response, 
capacitance ($C_Q$), and static Born charges ($\mathbf{Z}_{\kappa\alpha}^\text{stat}$).
Most of these quantities are straightforward to calculate by means of the existing implementation of density-functional perturbation theory (DFPT), which constitutes our general methodological framework. 
The only exception is $\mathbf{Z}^\text{stat}_{\kappa\alpha}$, which requires a long-wavelength expansion of the charge response to a phonon. We will discuss a convenient approach to perform such task by exploiting the ``$2n+1$'' theorem to the momentum derivative, in close analogy to the implementation of quadrupoles in insulators \cite{Royo2019}. 
As we shall see shortly, this strategy avoids the need for a numerical fitting of 
finite-$\mathbf{q}$ functions altogether, since it only involves calculating one additional response function at the zone center: the SCF response to a uniform scalar potential at fixed Fermi level, which we have also 
implemented in \textsc{abinit} \cite{Abinit_1}. Such calculation provides as a by-product the quantum capacitance via Eq. (\ref{eq:C_DFPT}). 
\subsection{Basics of DFPT}
We start by assuming a parametric dependence of the Hamiltonian on a small
parameter $\lambda$, which describes the external perturbation. 
By means of standard time-independent perturbation theory techniques, the physical properties of the system are expanded in powers of $\lambda$,
\begin{equation}\label{Eq_X_lambda}
X(\lambda)=\sum_n \frac{1}{n!}\lambda^n X^{(n)},\quad 
X^{(n)}=\frac{d X(\lambda)}{d\lambda^n}\Big|_{\lambda=0},
\end{equation}
where $X$ can be the Hamiltonian $\hat{H}$, the wave functions $\psi_{n\mathbf{k}}(\mathbf{r})$ or the 
energy $E$. (Notice that the expansion coefficients in 
Eq. (\ref{Eq_X_lambda}) are directly related to the
derivatives of $X$ with respect to $\lambda$; this convention 
differs from some earlier works \cite{PhysRevA.52.1096,Gonze1995,Gonze1997}.)
Due to the
variational character of the Kohn-Sham energy functional, the application of
perturbation theory results in a
``$2n+1$'' 
theorem, i.e., the knowledge of the $n$-th order wave functions 
are enough to compute ($2n + 1$)-th order energies, and even
derivatives of the energies acquire a variational 
character \cite{PhysRevB.39.13120,PhysRevA.52.1096}.

For our scopes, second-order energy derivatives are 
sufficient; therefore, we only need to access first-order
wave functions, $\psi_{n\mathbf{k,q}}^\lambda(\mathbf{r})=
e^{i(\mathbf{k+q})\cdot\mathbf{r}} u^\lambda_{n\mathbf{k,q}}(\mathbf{r})$.
To see how the machinery of DFPT facilitates the calculation of such derivatives,
one can start from the Kubo formula for the electronic contribution to the response,
\begin{equation}
\begin{split}
E^{\lambda_1^* \lambda_2}_{\bf q} = & \int_\text{BZ} [d^3 k] \sum_{ij} \frac{f_{j\bf k+q}-f_{i\bf k}}{\e_{j\bf k+q}-\e_{i\bf k}} \times
\\
&\qquad  \langle u_{i \bf k}|(H^{\lambda_1}_{\bf k,q})^\dagger| u_{j \bf k+q} \rangle 
\langle u_{j \bf k+q}| \hat{\mathcal{H}}^{\lambda_2}_{\bf k,q} |u_{i \bf k} \rangle.
\end{split}
\end{equation}
The DFPT approach consists in limiting the double summation over $ij$ to an \emph{active subspace} 
spanned by the lowest $M$ states.
If the remainder orbitals have vanishing population, their contribution can be written in the 
form
\begin{equation}
\Delta E^{\lambda_1^* \lambda_2}_{\bf q} =  2 \int_\text{BZ} [d^3 k] \sum_{i=1}^M f_{i\bf k} \langle u_{i \bf k}|(H^{\lambda_1}_{\bf k,q})^\dagger|  u^{\lambda_2}_{n\mathbf{k,q}}\rangle.
\end{equation}
The first-order wave functions, in turn, can be obtained from the following 
Sternheimer equation, as proposed by Baroni \textit{et al.} \cite{Baroni2001},
\begin{equation}\label{Eq_Stern_Baroni}
(\hat{H}_\mathbf{k+q}
+a\hat{P}_\mathbf{k+q}-\epsilon_{n\mathbf{k}})
\ket*{u_{n\mathbf{k,q}}^\lambda}=
-\hat{Q}_\mathbf{k+q}\mathcal{\hat{H}}^\lambda_\mathbf{k,q}\ket*{u_{n\mathbf{k}}},
\end{equation}
where $a$ is a constant with dimension of energy that ensures
that the left hand side of Eq.~(\ref{Eq_Stern_Baroni}) does not
become singular. 
(To simplify the notation,
we use $\ket*{u_{n\mathbf{k}}}=\ket*{u^{(0)}_{n\mathbf{k}}}$, 
$\epsilon_{n\mathbf{k}}=\epsilon_{n\mathbf{k}}^{(0)}$ and 
$\hat{H}_{\mathbf{k}}=\hat{H}^{(0)}_\mathbf{k}$.)
The operators
\begin{equation}
\hat{P}_\mathbf{k+q}=\sum_{m=1}^M \ket*{u_{m\mathbf{k}}}
\bra*{u_{m\mathbf{k}}}, \quad \hat{Q}_\mathbf{k+q}=1-\hat{P}_\mathbf{k+q}
\end{equation}
are projectors onto and out of the active subspace, respectively.

In insulators, $M$ is most conveniently set to the number of occupied states. 
Then, $\hat{P}_\mathbf{k+q}$ and $\hat{Q}_\mathbf{k+q}$ reduce 
to the valence- and conduction-band projectors, respectively, and the Kubo-like contribution to the second derivative vanishes identically.  
In metals, on the other hand, there is no clear distinction
between occupied and unoccupied manifolds, as the Fermi-Dirac distribution function
acquires fractional values around the Fermi level.
$M$ should be then set to a large enough value to include all partially populated states, in such a way that the occupancy of the excluded bands is negligibly small. \cite{Gonze2024,Dreyer2022,zabalo2024ensemble}
At this point, a further increase in the \emph{active subspace} dimension, $M$, shouldn't have any impact on 
any measurable quantity, e.g., 
the ground state energy or its derivatives with respect to $\lambda$.
Thus the parameter $M$ is, to a large extent, 
arbitrary; this freedom can be exploited to accelerate convergence.
\subsection{Long-wave DFPT}
It was recently demonstrated in Ref. \cite{Royo2019}
that the wave vector
$\mathbf{q}$ can be treated as an additional perturbation parameter of 
the energy functional, which allows the application of standard tools of DFPT and, 
in particular, the ``$2n+1$'' theorem, with the wave vector. 

Consider the mixed derivative of the energy
with respect to two external perturbations, $\lambda_1$ and $\lambda_2$,
\begin{equation}\label{Eq_E_con}
E_{\text{con},\mathbf{q}}^{\lambda_1\lambda_2}=\frac{d^2 E_\mathbf{q}}{d\lambda_1 d\lambda_2},
\end{equation}
which is a variational functional of the first-order wave functions, 
$\ket*{u^{\lambda_{1,2}}_{m\mathbf{k}}}$. (The stationary condition 
$\delta E_\mathbf{q}^{\lambda_1\lambda_2}/\delta \bra*{u^{\lambda_1}_{n\mathbf{k}}}=0$
gives rise to a Sternheimer equation analogous to that 
presented in Eq. (\ref{Eq_Stern_Baroni}) for the 
first-order wave functions $\ket*{u_{m\mathbf{k}}^{\lambda_2}}$.) 
$E_{\text{con},\mathbf{q}}^{\lambda_1\lambda_2}$ is
usually minimized under the constraint (con) that
$\bra*{u_{n\mathbf{k+q}}}\ket*{u^\lambda_{m\mathbf{k}}}=0$,
for all $m,n$ considered in the calculation; this is known as the parallel 
transport gauge.
Due to the inherent phase indeterminacy of the $\ket*{u_{m\mathbf{k+q}}}$
Bloch functions, 
taking the first $\mathbf{q}$ derivative of Eq. (\ref{Eq_E_con}) 
is not obvious.

To overcome this issue, in Ref. \cite{Royo2019} the linear response problem is reformulated as 
finding the variational minimum of a new \textit{unconstrained} second-order energy, 
$E_\mathbf{q}^{\lambda_1\lambda_2}$. 
(Of course, the variational minimum of both the 
constrained and the unconstrained second-order energy as 
proposed in Ref. \cite{Royo2019}
lead to the exact same solution.)
The key feature of the unconstrained energy functional as defined 
in Ref. \cite{Royo2019} is that
the only objects that explicitly depend on $\mathbf{q}$ are the operators, which
are well defined mathematical objects that do not suffer from the phase indeterminacy
of the Bloch functions.
It is then feasible to obtain an analytical long-wavelength expansion in powers 
of $\mathbf{q}$ of the second-order energy, which can be pushed to any order in the wave vector. 
At first order in 
$\mathbf{q}$, the ``$2n+1$'' theorem reduces to the Hellmann-Feynman theorem,
\begin{equation}
 E_\gamma^{\lambda_1\lambda_2}=
 \frac{d E_\mathbf{q}^{\lambda_1\lambda_2}}{d q_\gamma}\bigg|_\mathbf{q=0}=
 \frac{\partial E_\mathbf{q}^{\lambda_1\lambda_2}}{\partial q_\gamma}\bigg|_\mathbf{q=0},
\end{equation}
which states that the total derivative in $\mathbf{q}$ coincides with 
its partial derivative, meaning that
first-order wave functions, $\ket*{u^\lambda_{m\mathbf{k,q}}}$, are excluded from differentiation \cite{Royo2019}. 
As a consequence, the sole knowledge of wave function responses to uniform ($\mathbf{q=0}$) 
field perturbations, 
$\ket*{u^\lambda_{m\mathbf{k}}}=\ket*{u^\lambda_{m\mathbf{k,q=0}}}$,
is enough to access first-order spatial dispersion coefficients.
Successful applications of this method 
have been demonstrated in several contexts,
including dynamical quadrupoles \cite{Royo2019}, 
flexoelectricity \cite{Royo2022} or 
natural optical activity \cite{PhysRevLett.131.086902}.

Note that the long-wavelength DFPT methodology of Ref. \cite{Royo2019} has been 
recently generalized to metals in
Ref. \cite{zabalo2024ensemble}. In the next subsection we will apply this 
method to obtain an analytical long-wavelength expansion of 
the charge density response to a phonon, which gives access to 
$\textbf{Z}^{\text{stat}}_{\kappa\alpha}$.
\subsection{Direct calculation of $\textbf{Z}^{\text{stat}}_{\kappa\alpha}$ via analytic differentiation in \textbf{q}}\label{sec:analytic_q}
Our starting point is the first $\mathbf{q}$ derivative of 
Eq. (\ref{eq:chi_rho_met_b}),
\begin{equation}\label{Eq_Z_stat_rho}
Z^{\text{stat},\beta}_{\kappa\alpha}=i\Omega 
\frac{\partial \rho^\mathbf{q}_{\kappa\alpha}}{\partial q_\beta}\Big|_\mathbf{q=0}
=i\Omega \rho_{\kappa\alpha}^{(1,\beta)},
\end{equation}
where the charge density response to a phonon at linear order in $\mathbf{q}$,
$\rho^{(1,\beta)}_{\kappa\alpha}$, contains both ionic and electronic (el) contributions.
(For simplicity, we have dropped the ``SR'' label.)
The ionic contribution to Eq. (\ref{Eq_Z_stat_rho}) is proportional 
to the pseudocharge of the ions and,
therefore, trivial. 
Regarding the electronic contribution, 
in order to take advantage of the analytical long-wave DFPT method, 
we express the 
charge density response to a phonon as a second derivative of the energy with respect to 
a scalar perturbation ($\varphi$) and an 
atomic displacement ($\tau_{\kappa\alpha}$), and proceed to take its first 
$\mathbf{q}$ derivative by exploiting the ``$2n+1$'' theorem,
as described in Refs. \cite{Royo2019,zabalo2024ensemble},
\begin{equation}
\Omega\rho^{(1,\beta)}_{\text{el},\kappa\alpha}=
\frac{\partial E_\mathbf{q}^{\varphi\tau_{\kappa\alpha}}}
{\partial q_\beta}\Big|_\mathbf{q=0}=
E_\beta^{\varphi\tau_{\kappa\alpha}}.
\end{equation}
The latter immediately leads to the following expression for the 
\textit{static Born charges},
\begin{equation}
Z^{\text{stat},\beta}_{\kappa\alpha}=
-\text{Im}\,E_\beta^{\varphi\tau_{\kappa\alpha}}
+\delta_{\alpha\beta}Z_\kappa,
\end{equation}
where $Z_\kappa$ is the pseudocharge of ion $\kappa$, and 
$E_\beta^{\varphi\tau_{\kappa\alpha}}$ can be obtained following the
guidelines of Ref. \cite{zabalo2024ensemble}. In particular, we find
\begin{equation}\label{Eq_E_lw}
\begin{split}
E_\beta^{\varphi\tau_{\kappa\alpha}}=&\int_\text{BZ}[d^3k]
\left( 2\bar{E}_\beta^{\varphi\tau_{\kappa\alpha}} 
+\Delta E_\beta^{\varphi\tau_{\kappa\alpha}}\right)\\
&+\int_\Omega\int \rho^{\varphi}(\mathbf{r})
K_\beta(\mathbf{r},\mathbf{r'}) 
\rho^{\tau_{\kappa\alpha}}(\mathbf{r'})
\, d^3r \, d^3r',
\end{split}
\end{equation}
where $K_\beta(\mathbf{r,r'})$ represents the first $\mathbf{q}$ gradient of the 
Hartree and exchange-correlation kernel \cite{Royo2019}, and 
$\bar{E}_\beta^{\varphi\tau_{\kappa\alpha}}$ 
and $\Delta E_\beta^{\varphi\tau_{\kappa\alpha}}$ are, respectively, the 
\textit{wave function} and \textit{occupation} contributions as
defined in Ref. \cite{zabalo2024ensemble}. For our particular case,  
\begin{widetext}
\begin{equation}\label{Eq_wf_term}
\begin{split}
\bar{E}_\beta^{\varphi\tau_{\kappa\alpha}}=&
\sum_{m=1}^M f_{m\mathbf{k}} \left(\bra*{u^\varphi_{m\mathbf{k}}}
\hat{H}^{k_\beta}_\mathbf{k}\ket*{u^{\tau_{\kappa\alpha}}_{m\mathbf{k}}}
+\bra*{u^\varphi_{m\mathbf{k}}}\hat{H}_{\mathbf{k},\beta}^{\tau_{\kappa\alpha}}
\ket*{u_{m\mathbf{k}}}
\right)\\
&-\sum_{m,n=1}^M f_{m\mathbf{k}}
\left(
\bra*{u_{m\mathbf{k}}}\hat{\mathcal{H}}^\varphi
\ket*{u_{n\mathbf{k}}}
\bra*{u_{n\mathbf{k}}^{k_\beta}}\ket*{u_{m\mathbf{k}}^{\tau_{\kappa\alpha}}}
+\bra*{u^\varphi_{m\mathbf{k}}}\ket*{u^{k_\beta}_{n\mathbf{k}}}
\bra*{u_{n\mathbf{k}}}
\hat{\mathcal{H}}_\mathbf{k}^{\tau_{\kappa\alpha}}\ket*{u_{m\mathbf{k}}}
\right)
\end{split}
\end{equation}
and 
\begin{equation}\label{Eq_occ_term}
\begin{split}
\Delta E_\beta^{\varphi\tau_{\kappa\alpha}}=&\sum_{m,n=1}^M
\bar{f}_{mn\mathbf{k}} 
\left(
 \bra*{u_{m\mathbf{k}}} 
 \hat{\mathcal{H}}^\varphi
 \ket*{u_{n\mathbf{k}}}
 \bra*{u_{n\mathbf{k}}}\hat{\mathcal{H}}^{\tau_{\kappa\alpha}}_{\mathbf{k},\beta}
 \ket*{u_{m\mathbf{k}}}
\right)\\
&+\sum_{m,n=1}^M \bar{f}_{mn\mathbf{k}}
\left(
 \bra*{u_{m\mathbf{k}}}
 \hat{\mathcal{H}}^\varphi
 \ket*{u_{n\mathbf{k}}^{k_\beta}}\bra*{u_{n\mathbf{k}}}
 \hat{\mathcal{H}}_\mathbf{k}^{\tau_{\kappa\alpha}}\ket*{u_{m\mathbf{k}}}
+\bra*{u_{m\mathbf{k}}}
 \hat{\mathcal{H}}^\varphi
 \ket*{u_{m\mathbf{k}}}
 \bra*{u^{k_\beta}_{n\mathbf{k}}}
 \hat{\mathcal{H}}_\mathbf{k}^{\tau_{\kappa\alpha}}
 \ket*{u_{m\mathbf{k}}}
\right) \\
&+\sum_{m,n,l=1}^M \mathcal{F}_{mnl\mathbf{k}}
\bra*{u_{m\mathbf{k}}}
\hat{\mathcal{H}}^\varphi
\ket*{u_{n\mathbf{k}}}
\bra*{u_{n\mathbf{k}}}\hat{H}^{k_\beta}_\mathbf{k}\ket*{u_{l\mathbf{k}}}
\bra*{u_{l\mathbf{k}}}\hat{\mathcal{H}}_\mathbf{k}^{\tau_{\kappa\alpha}}
\ket*{u_{m\mathbf{k}}},
\end{split}
\end{equation}
\end{widetext}
(The derivative of the external scalar potential $H^\varphi_{\bf q}$ with respect to $q_\beta$ vanishes, 
hence the apparent lack of symmetry of some terms with respect to the two perturbation parameters $\tau_{\kappa\alpha}$ 
and $\varphi$.)
In the following, we shall explain the new terms appearing 
in the last two equations.
Regarding operators, we have defined
\begin{equation}
\hat{H}_\mathbf{k}^{k_\beta}=\frac{\partial\hat{H}_\mathbf{k+q}}
{\partial q_\beta}\Big|_\mathbf{q=0}\quad\text{and}\quad \hat{H}^{\tau_{\kappa\alpha}}_{\mathbf{k},\beta}=
\frac{\partial\hat{H}_\mathbf{k,q}^{\tau_{\kappa\alpha}}}
{\partial q_\beta}\Big|_\mathbf{q=0},
\end{equation}
where $\hat{H}_\mathbf{k}^{k_\beta}$ is commonly known as the velocity operator.
Apart from that, we have introduced the following new symbols 
\cite{Stengel2000,zabalo2024ensemble}, 
\begin{equation}\label{Eq_bar_f}
\bar{f}_{mn\mathbf{k}}=
\begin{cases}
\frac{f_{m\mathbf{k}}-f_{n\mathbf{k}}}
{\epsilon_{m\mathbf{k}}-\epsilon_{n\mathbf{k}}}&,\quad\text{if } \epsilon_{m\mathbf{k}}\neq \epsilon_{n\mathbf{k}},\\[10pt]
\frac{f'_{m\mathbf{k}} + f'_{n\mathbf{k}}}{2}
&, \quad\text{if } \epsilon_{m\mathbf{k}}=\epsilon_{n\mathbf{k}}.
\end{cases}
\end{equation}
and
\begin{equation}
\mathcal{F}_{mnl\mathbf{k}}=
\begin{cases}
\frac{\bar{f}_{mn\mathbf{k}}-\bar{f}_{ml\mathbf{k}}}
{\epsilon_{n\mathbf{k}}-\epsilon_{l\mathbf{k}}}&,\quad\text{if }
\epsilon_{n\mathbf{k}}\neq \epsilon_{l\mathbf{k}},\\[10pt]
\frac{f''_{m\mathbf{k}}+f''_{n\mathbf{k}}+f''_{l\mathbf{k}}}{6}&,
\quad\text{if } 
\epsilon_{m\mathbf{k}}=\epsilon_{n\mathbf{k}}=\epsilon_{l\mathbf{k}},
\end{cases}
\end{equation}
where $f''_{m\mathbf{k}}=\partial^2 f_{m\mathbf{k}}/\partial\epsilon^2_{m\mathbf{k}}$ 
is the second energy derivative of the
occupation function and  
$\mathcal{F}_{mnl\mathbf{k}}$ remains invariant under any permutation of the three band indices $m,n,l$. (The need for a numerical tolerance to determine 
$\epsilon_{m\mathbf{k}}=\epsilon_{n\mathbf{k}}$ explains the 
symmetrization of the indices in the second line of 
Eq. (\ref{Eq_bar_f}); the same argument applies for the definition of 
$\mathcal{F}_{mnl\mathbf{k}}$ when
$\epsilon_{m\mathbf{k}}=\epsilon_{n\mathbf{k}}=\epsilon_{l\mathbf{k}}$.)
All the first-order wave functions appearing in Eq. (\ref{Eq_wf_term}) 
and (\ref{Eq_occ_term}), 
$\ket*{u^{\tau_{\kappa\alpha}}_{m\mathbf{k}}}$, 
$\ket*{u^{k_\beta}_{m\mathbf{k}}}$ and $\ket*{u^\varphi_{m\mathbf{k}}}$, can be obtained
from a Sternheimer equation, as defined in Eq.~(\ref{Eq_Stern_Baroni}),
upon substitution of their corresponding perturbing Hamiltonian, 
$\hat{\mathcal{H}}^\lambda_\mathbf{k,q}$. 
Note that the so-called Fermi level shifts must be switched off in the linear 
response calculations, consistent with the requirement that the derivatives must be taken at
fixed Fermi energy.

An additional detail that requires further attention is that,
in order for the methodology of Ref.~\citenum{zabalo2024ensemble} 
to be applicable,
the highest state in energy considered in the calculation, $M$,
needs to be separated in energy to that of $M+1$, and this needs to be fulfilled 
for all $\mathbf{k}$ points in the Brillouin Zone. In other words, 
the active subspace ($\mathcal{M}$)  
needs to form an isolated subset of bands such that 
$\epsilon_{M\mathbf{k}}\neq \epsilon_{M+1\mathbf{k}}, \forall \mathbf{k}\in 
\text{BZ}$. (A more detailed discussion of this topic can be found 
in Section III.B of Ref.~\cite{zabalo2024ensemble}.)
While certainly possible, generalizing 
Eq. (\ref{Eq_E_lw}) to a case in which
such degeneracies are allowed goes beyond the scopes of the present work.
There are, however, fortunate cases where a well defined energy gap exists
for all $\mathbf{k}$ points in the 
electronic band structure. 
As we shall see in the following section, this is in fact the case with LiOsO$_3$
and SrTiO$_3$. Unfortunately, we were unable to find 
an isolated group of bands for the other test cases, i.e., 
aluminum and cooper; however, we can still rely on finite $\mathbf{q}$ calculations to demonstrate our claims.

\section{Results \label{sec:res}}

We now turn to our numerical results aimed at demonstrating the physics discussed in Sec.~\ref{sec:form}. In particular, we present calculations for four key quantities discussed above: the static charge response to a phonon \Zstat{}, the quantum capacitance $C_Q$, the Fermi-level deformation potential $\mu^{\eta_{\alpha\beta}}\equiv \partial \mu/\partial \eta_{\alpha\beta}$, and the charge response to a uniform strain $N^{\eta_{\alpha\beta}}\equiv\partial N/\partial \eta_{\alpha\beta}$. For \Zstat{}, we validate the methodology described in Sec.~\ref{sec:analytic_q}, testing against the conventional approach involving numerical differentiation in $q$. For $C_Q$, we demonstrate the importance of self-consistent fields for obtaining an accurate value. With these quantities in hand, we prove the sum rule of Eq.~(\ref{eq:sum_rule2}), and relate the product of the deformation potential and quantum capacitance to $N^{\eta_{\alpha\beta}}$. Finally, we demonstrate the reference potential dependence of \Zstat{} 
and $\mu^{\eta_{\alpha\beta}}$ discussed in Sec.~\ref{sec:ref_dep}.

We consider several test case materials. The first two are simple metals: face-centered cubic (FCC) aluminum, and copper. We also present results on the polar metal  \loo{} \cite{shi2013ferroelectric}, which has a more nontrivial sublattice sum, since there are multiple different atoms per formula unit; for simplicity, we shall
restrict our analysis to the cubic phase with $Pm\overline{3}m$ symmetry. Finally, we explore doping of the cubic $Pm\overline{3}m$ structure of \sto{}, which is an insulator in its ground state, but experimentally has interesting properties when lightly doped $n$-type \cite{Schooley1965,Huang2017,Collignon2019}. We will denote the doped material as \sto{}$(n)$ where $n$ is the charge in $e$ per unit cell added or removed (negative values indicate extra electrons).

Since all materials considered have cubic symmetry, it is sufficient to consider one strain/atomic displacement direction, so to simplify the notation we take $Z_{\kappa x}^{\text{stat},x}=Z_{\kappa y}^{\text{stat},y}=Z_{\kappa z}^{\text{stat},z}\equiv Z_\kappa^{\text{stat}}$ and $\mu^{\eta_{xx}}=\mu^{\eta_{yy}}=\mu^{\eta_{zz}}\equiv \mu^\eta$ (and similarly for $N^\eta$).

\begin{table*}
\caption{Quantities discussed in Sec.~\ref{sec:form} for Al, Cu, \loo{}, and \sto($-0.05$) calculated wither with the point charge of local psudopotential choice for the reference potential. $C_Q$ ($C_0$) is the quantum capacitance with (without) self-consistent fields, $\mu^\eta$ is the Fermi level deformation potential at fixed charge, $N^\eta$ is the charge response to a uniform strain at fixed Fermi level, and \Zstatcub{} is the first-order charge response to a phonon. A $40\times40\times40$ $k$-mesh is used for Cu and Al, and $16\times16\times16$ for \loo{} and \sto($-0.05$).  \label{tab:sum_rule}}
\begin{ruledtabular}
	\begin{tabular}{c|cccccc} 
		 &reference potential&$C_0$ &$C_Q$ & $\mu^\eta$ & $N^\eta$ 
		& $\sum_\kappa Z_\kappa^{\text{stat}}$ \\\hline
		\multirow{ 2}{*}{Al} &pc & \multirow{ 2}{*}{10.715} & \multirow{ 2}{*}{16.485} & $-0.391$ & 6.441 & 6.448  \\
		                 &loc &  &  & $-0.385$ & 6.352 & 6.360 \\\hline
\multirow{ 2}{*}{Cu} & pc   &\multirow{ 2}{*}{7.941} &\multirow{ 2}{*}{8.002} &$-0.895$ &7.165 &7.158  \\
 & loc & & & $-0.599$ & 4.791 & 4.787 \\\hline
 		\multirow{ 2}{*}{\loo{}} &pc & \multirow{ 2}{*}{67.623} &\multirow{ 2}{*}{21.893} &$-0.687$ & 15.035 &15.031  \\
		 &loc & & & $-0.563$ & 12.329 & 12.325\\\hline
\multirow{ 2}{*}{\sto{}$(-0.05)$} & pc   & \multirow{ 2}{*}{12.202} & \multirow{ 2}{*}{9.557} & $-0.529$ & 5.059 & 5.060  \\
 & loc &   & &$-0.307$ &2.933 &2.933  \\
 \end{tabular}
\end{ruledtabular}	
\end{table*}

\subsection{Computational parameters}

All calculations were carried out using the {\sc abinit} code \cite{Abinit_1}, including the DFPT capabilities \cite{Gonze1997,GonzeLee1997} for phonon and strain \cite{Hamann2005} perturbations and the {\sc longwave} module \cite{Royo2019} for dispersion properties (i.e., analytic \textbf{q} derivatives). For the exchange-correlation functional, the local density approximation (LDA) was used with the Perdew-Wang parametrization \cite{Perdew1992}. For the primitive one-atom cells of Al and Cu, 
we use 
cell parameters of $a=7.520$ bohr and $a=6.645$ bohr, 
respectively.
For our calculations on 
LiOsO$_3$ and SrTiO$_3$, we employed a five-atom 
primitive cell with lattice constants of $a=7.130$ bohr and $a=7.288$ bohr. 
Structures were relaxed
ensuring
that the maximum stress component 
did not exceed $5\cross 10^{-9}$ Ha/bohr$^3$ in all cases.
To deal with the metallic occupations, we use Gaussian smearing with a broadening of $0.01$ Ha. Optimized Vanderbilt norm-conserving pseudopotentials \cite{Hamann2013} were used, generated from the {\sc pseudo-dojo} \cite{Setten2018} inputs, however with nonlinear exchange-correlation core corrections turned off to be compatible with the {\sc longwave} module of {\sc abinit}. Plane wave cutoffs of 20, 50, 60, and 60 Ha were used for Al, Cu, \loo{}, and \sto{}, respectively. We find that the number of $k$-points used to sample the Brillouin zone is a key convergence parameter and will be discussed below on a case-by-case basis.

In Table~\ref{tab:sum_rule} we give an overview of our main results for the quantum capacitance, the uniform strain 
response and the sum of the static Born charges (for \sto{} we focus on a doping of 0.05 $e/$u.c. as a representative case, 
where u.c. stands for unit cell).
To demonstrate the role played by the reference potential, we report values calculated with both of the choices discussed in Sec.~\ref{sec:ref_dep}, i.e., the Hartree potential plus the local part of the pseudopotentials (which we will refer to as ``loc'') or the Hartree potential plus compensating point charges (which we will refer to as ``pc'').
Comparing the last two columns of Table~\ref{tab:sum_rule}, we see immediately that the sum rule [Eq.~(\ref{eq:sum_rule2})] is satisfied in all materials for both reference potential choices up to an excellent numerical accuracy.
In the following we provide a detailed discussion of these results for each of the materials in our test set.

\subsection{Elemental metals: Al and Cu}\label{sec_el_metals}

\begin{figure}[t!]
	\includegraphics[width=1\linewidth]{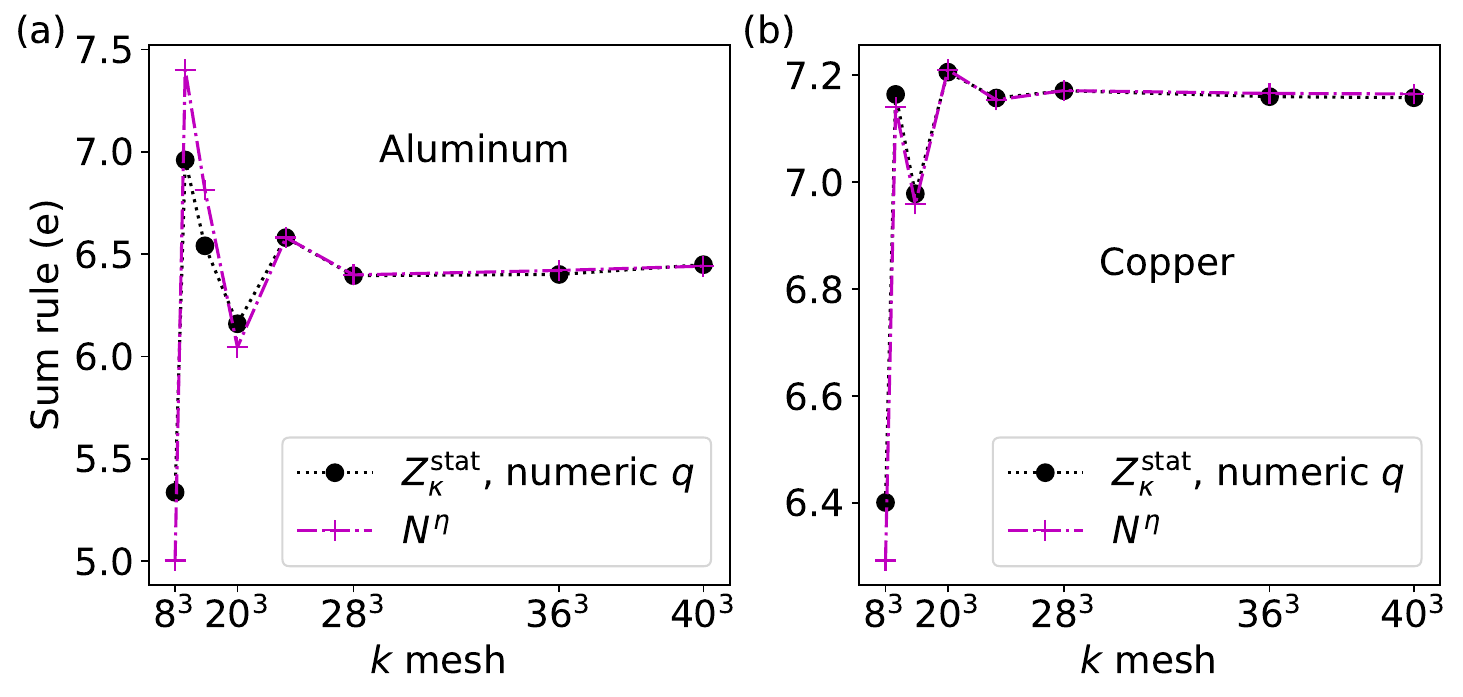}
	\caption{Numerical results demonstrating the sum rule of \eq{eq:sum_rule2} for (a) Al and (b) Cu. Black circles are the sublattice sum of the first-order density response calculated via a numerical $q$ derivative,  
    while the magenta +'s are the charge response to a strain at fixed Fermi level. The point charge reference potential is used.}
	\label{fig:sum_rules_cu_al}
\end{figure}

We begin with the results for elemental metals Al and Cu. In Fig.~\ref{fig:sum_rules_cu_al} we show the convergence of \Zstatcub{} \footnote{Note that for these materials we calculate \Zstatcub{} via numerical differentiation in $q$: the absence of energy gaps in the conduction bands makes the current implementation of the analytical long-wave approach (Sec.~\ref{sec:analytic_q}) unsuitable.} and $N^\eta$ versus $k$-mesh resolution. 
Note that a rather dense mesh is required to achieve convergence, possibly due to the highly dispersive states near the Fermi surfaces of these metals. Nevertheless, the sum rule appears to be already accurate at significantly coarser meshes, since $N^\eta$ and $Z_\kappa^{\text{stat}}$ appear to behave similarly with increasing resolution.  
As we anticipated in Section~\ref{sec:sum_rule}, $N^\eta$ can be obtained in two different ways. One can either perform a strain perturbation at fixed Fermi level \footnote{This nonstandard constraint is possible in {\sc abinit} via the \texttt{frzfermi} flag.} and determine the first-order charge, or one can calculate the Fermi-level deformation potential at fixed particle number and multiply the result by $C_{\rm Q}$ [see Eq.~(\ref{eq:sum_rule2})]. We have confirmed numerically that the two methods yield the same result for our materials test set. For example, for Cu and Al with a $40\times40\times40$ $k$-mesh, the values agree to more than six decimal places. 

As expected, the quantum capacitance does not depend on the choice of reference potential. Comparing Al and Cu, we see qualitatively different behavior when adding electron-electron interactions via SCF: for Al, $C_Q$ is significantly larger than $C_0$, while there is almost no change for Cu. We can shed further light on this difference by comparing the physics of the quantum capacitance in these nearly-free electron metals to the long-established theory of the uniform electron gas (UEG) \cite{Mahan2013}.
In the latter, the noninteracting (Lindhard) capacitance is (see Appendix~\ref{app:feg}) a simple function of the density $n$,
\begin{equation}
\frac{C^{\text{UEG}}_{0}}{\Omega} = \frac{m_e}{\pi^2 \hbar^2} (3\pi^2 n)^{1/3},
\end{equation}
where $m_e$ is the electron mass.
Regarding the SCF corrections, note that the Hartree local fields vanish in the uniform limit; within the range of densities relevant for real materials, the XC contribution is dominated by exchange, 
which yields a contribution to the inverse capacitance of
\begin{equation}
 \Omega (C^{\text{UEG}}_x)^{-1}  = - \frac{1}{3} \left( \frac{3}{\pi} \right)^{\frac{1}{3}}  n^{-\frac{2}{3}}.
\end{equation}

By plugging in the calculated values for the equilibrium volume and number of free electrons per cell 
(respectively three and one for Al and Cu) in the above formulas, we obtain the values reported in Table~\ref{tab:C_UEG}. 
The case of Al is in remarkable agreement with the results of our first-principles calculations, especially once we take into account the effective mass of 1.06 
This outcome indicates that Al is indeed a nearly-free electron metal, with spatial fluctuations in the density and effective background potential that are largely irrelevant to its static screening properties.
To corroborate this point further, we have repeated the first-principles computation of the quantum capacitance by suppressing the XC fields completely. 
The result, within 1\% of $C_0$, confirms our hypothesis that the Hartree local fields have a negligible impact on $C_{\rm Q}$.
The calculated results for FCC Cu, on the other hand, are in marked disagreement with the predictions of the uniform gas model: the latter underestimates $C_0$ and severely overestimates the SCF corrections. Further inspection show that the negative XC contribution to $C_{\rm Q}$ cancels out almost exactly with the positive contribution of the Hartree local fields, which explains the surprising result (Table I) that $C_0\simeq C_{\rm Q}$ in this material.   
Thus, while often regarded as a nearly-free-electron metal, our results show that in terms of its quantum capacitance, Cu does not really behave as such.

We now move to the discussion of the deformation potential, $\mu^\eta$, and the static charges, \Zstatcub{}. 
As anticipated earlier, both quantities show a clear dependence on the choice of reference potential; nevertheless, the sum rule is always satisfied, provided that a consistent convention is used in calculating $\mu^\eta$ and \Zstatcub{}.
Based on Eq.~(\ref{eq:ref_dep}) we can predict the difference of \Zstatcub{} between the pc and loc references just by knowing
the properties of the pseudopotential.
Using Eq.~(\ref{eq:ref_dep}), we predict $\Delta Z_{\text{Al}}^{\text{stat}}=0.088$ eV and $\Delta Z_{\text{Cu}}^{\text{stat}}=2.373$ eV, which is in excellent agreement with 0.089 and 2.370 eV (respectively) from Table~\ref{tab:sum_rule}.
We note that our value for $Z_{\text{Al}}^{\text{stat}}$ is in very good agreement with the value of 6.482 $e$ calculated in Ref.~\citenum{Marchese2023} by using the pc reference. 
In a more recent work \cite{Macheda2024}, a value of 6.80 $e$ was reported, and  $\Delta Z_{\text{Al}}^{\text{stat}}$ found to be 0.169 eV. This value was consistent with an analogous equation to Eq.~(\ref{eq:ref_dep}) given in Ref.~\citenum{Macheda2024}, indicating that indeed this quantity is dependent on both the reference potential chosen, and the properties of the pseudopotential.

As pointed out previously \cite{Marchese2023}, $Z_{\text{Al}}^{\text{stat}}$ differs significantly from the nonadiabatic or dynamical Born charge of Al,
$Z_{\text{Al}}^{\text{dyn}}$, first reported as 1.969 $e$ in Ref.~\citenum{Dreyer2022}, and later confirmed by Refs.~\citenum{Wang2022,Marchese2023} (reporting 2.09 and 1.987 $e$, respectively). 
An obvious question to ask is why \Zstatcub{} has such an anomalously large value in both Al and Cu, especially when compared with its dynamical counterpart.
Given that \Zstatcub{} coincides with $N^\eta$ in these elemental materials, one can seek an answer by analyzing the physical effects contributing to $\mu^\eta$, which is related to $N^\eta$ via a factor of $-C_{\rm Q}$.
Within the uniform electron gas model and with the pc reference, the Fermi level shift is given by (see Appendix \add{\ref{app:feg}})
\begin{equation}
\label{eq:mu_hom}
\mu^\eta = -\frac{N_e}{C_{\rm Q}},
\end{equation}
where $N_e$ is the number of free electrons per cell.
Therefore, this model yields \Zstatcub{}$=N_e$, i.e., the static charges coincide with the nominal external dipole
\footnote{The physical interpretation of \Zstatcub{} for the uniform electron gas model can be thought of via decomposing the compensating jellium into a lattice of broad gaussian charges, and considering the SCF response from the displacement of said charges.}
and with the dynamical ones (that \Zdyncub{}=$N_e$ in a uniform electron gas is a direct consequence of the $f$-sum rule~\cite{Resta2018}).
The calculated values of $\mu^\eta$, however, drastically depart from the prediction of this simplified model. Even in Al, whose capacitance is in excellent agreement with that of a uniform gas with the same density, $\mu^\eta$ is more than twice as large as the hypothetical value given by Eq.~(\ref{eq:mu_hom}). 
These deviations must originate from the inhomogeneities in the density response to a lattice distortion, which always exist in a real crystal. 
This observation is consistent with the exact relation, 
established in Eq.~(\ref{eq:delta_z}), between \Zstatcub{} and the octupolar moment of the screened charge induced by an atomic displacement: octupoles are indeed sensitive to the microscopic form factor of the displaced electron density.
This behavior is in stark contrast with that of \Zdyncub{}, whose sum is related to the inertia of the free carriers \cite{Resta2018,Dreyer2022}, and hence more clearly linked to macroscopic observables of the crystal.

\begin{table}
\caption{Quantum capacitance for the uniform electron gas model using the parameters for Al and Cu. $C^{\text{UEG}}_0$ is the noninteracting  case, while $C^{\text{UEG}}_{\rm Q}$ includes the exchange contribution. The second row incorporates the nonunity effective mass of Al.  \label{tab:C_UEG}}
\begin{ruledtabular}
    \begin{tabular}{l|cc}
     & $C^{\text{UEG}}_0$ & $C^{\text{UEG}}_{\rm Q}$   \\
     \hline
 Al($m^*=1$)       & 10.15   & 15.33 \\
 Al($m^*=1.06$)    & 10.75   & 16.76 \\
  Cu               &  5.49   &  9.65
    \end{tabular}
\end{ruledtabular}
\end{table}

\subsection{``Polar'' metal {\loo{}}\label{sec:loo}}

\begin{figure}
	\includegraphics[width=1\linewidth]{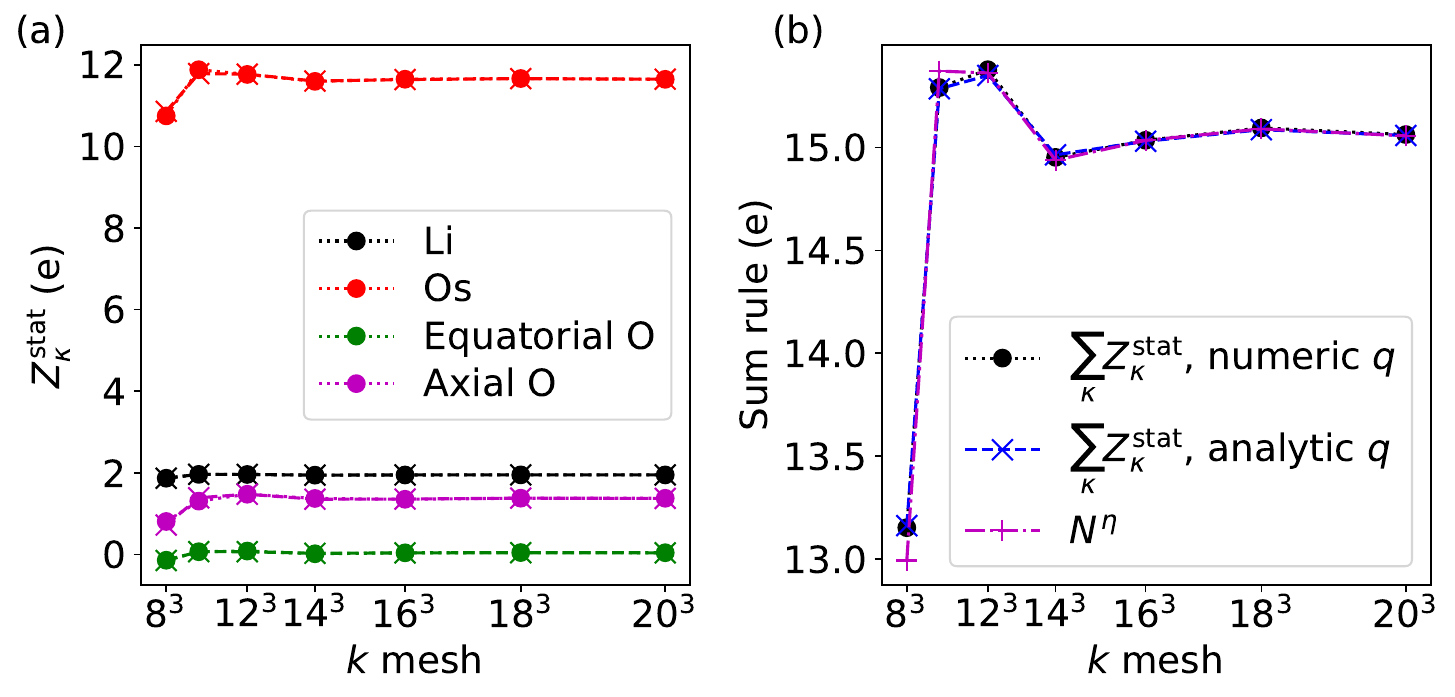}
	\caption{(a) \Zstatcub{}for the sublattices in cubic \loo{} versus $k$-mesh. Circles are calculated with the conventional methodology based on numerical $q$ derivatives, $\times$'s are calculated via  analytical $q$ derivatives described in Sec.~\ref{sec:analytic_q}. (b) numerical results demonstrating the sum rule. All calculations are using the H+pc reference potential. }
	\label{fig:loo}
\end{figure}

We now move to \loo{}, a material that differs substantially from the elemental metals of the previous section. First, like other metallic $AB$O$_3$ perovskites, it is a compound crystal with formally charged anions and cations. Second, its Fermi level falls within the $d$-electron band of the $B$-site cation (in this case, Os), in contrast with the $sp$ nature of the carriers in Al and Cu.
These characteristics, together with its popularity as the first experimental demonstration of a ``polar'' or ``ferroelectric'' metal \cite{shi2013ferroelectric}, make \loo{} an ideal candidate to complement our test set. Also, this material  exhibits energy gaps in the unoccupied part of the Kohn-Sham spectrum, and therefore it allows for a direct test of our computational methodology based on the ``$2n+1$'' theorem.

In Fig.~\ref{fig:loo} we present the results of our calculations on \loo{} using the pc reference. Panel (a) shows \Zstatcub{} for the different sublattices versus $k$-mesh. We can see that convergence requires significantly less $k$ points compared to the elemental metals, due to the much smaller dispersion of the bands around the Fermi surface in \loo{}. We show calculations using the conventional numerical $q$ derivative as well as the analytic $q$ derivative methodology discussed in Sec.~\ref{sec:analytic_q}; the points lie on top of each other for both the individual \Zstatcub{} [Fig.~\ref{fig:loo}(a)] and the sublattice sum [Fig.~\ref{fig:loo}(b)], validating our methodology. 
It is interesting to note that the calculated values of \Zstatcub{} are \emph{positive} for all sublattices, which is surprising especially for the O atoms, given their formal valence of $-2e$. Such an outcome is likely connected to the relation between the \Zstatcub{} and the screened dynamical octupoles: the $(xxx)$ components of the latter tend to be large and negative in compounds with predominantly ionic bonding, which is consistent with a positive contribution to the \Zstatcub{}. 
As a result of this fact, \loo{} displays a massive violation of the acoustic sum rule, with a total in excess of $15e$ (see Table I). To put this value in perspective, note that the violation of the \emph{dynamical} ASR in doped SrTiO$_3$ peaks at a maximum of $1e$ for a comparable population of the $t_{2g}$ band of Ti \cite{Dreyer2022}.

Next, we test the impact of the choice of the reference potential on our results. In Table~\ref{tab:loo_becs} we report
the values for \Zstatcub{} in \loo{}, calculated with 
both the pc and loc convention. As with the elemental metals, the difference between \Zstatcub{}(loc) and \Zstatcub{}(pc) 
nicely matches the analytical prediction of Eq.~(\ref{eq:ref_dep}). 
In Fig.~\ref{fig:loo}(b), we compare the sublattice sum of \Zstatcub{} with $N^\eta$, demonstrating that the sum rule is also satisfied to high accuracy. 
The  quantities entering the sum rule are reported in Table~\ref{tab:sum_rule} for 
both reference potentials. Again, the values of $\mu^\eta$ and \Zstatcub{} are different for different reference potentials, while the sum rule is satisfied for both choices. Also, the second equality of Eq.~(\ref{eq:sum_rule2}), i.e., $-\mu^\eta C_Q=N^\eta$ is accurately satisfied for \loo{}.

\begin{table}
\caption{Dependence on electrostatic potential reference of $Z_\kappa^{\text{stat}}$ for the different sublattices of \loo{}. ``Equat.~O'' (``axial O'') refers to the O atoms whose Os-O bonds are perpendicular (parallel) to the displacement direction. Calculations are performed using a $16\times16\times16$ $k$-mesh. ``diff.'' is the difference between the values for the different reference potentials and $\Delta Z^{\text{stat}}$ is the predicted difference between the reference potentials from Eq.~(\ref{eq:ref_dep}).  \label{tab:loo_becs}}
\begin{ruledtabular}
    \begin{tabular}{c|cccc} 
    ref. pot.&Li&Os&Equat. O&Axial O\\\hline
pc &$1.951$&$11.637$&$0.041$&$1.360$\\
loc& $1.793$&$10.195$&$-0.327$&$0.991$\\
\hline
diff. & 0.158 & 1.442  &0.368  & 0.369\\
$\Delta Z^{\text{stat}}$ & 0.158 & 1.442 &  0.369 &  0.369 \\
\end{tabular}
\end{ruledtabular}
\end{table}

Finally, we comment briefly on our results for the capacitance, which presents an interesting contrast to the elemental metals.
The non-SCF capacitance, $C_0$, is significantly larger, which is easily explained by observing that the density of states in the $d$-electron bands of Os is much higher than in the dispersive $sp$ bands of Cu and Al. 
Moreover, switching on the Hartree/XC fields results in a strong suppression of the capacitance, very much unlike the enhancement that we find in Al. 
Upon inspection, we find that this suppression is dominated by the behavior of the Hartree potential, which is sensitive to the strong Coulomb interactions within the localized $d$ orbitals. 
The XC potential is negative, as expected, but a little less than half the 
magnitude than the Hartree term, which explains the overall suppression.
We note that the localized nature of the $d$ electrons means that beyond semilocal DFT functionals are often necessary for an accurate description, e.g., DFT+$U$ or hybrid functionals; it would be interesting to test the effects of these methods on the results discussed here.

\subsection{Doped {\sto{}}}
Our last test case is a semiconductor, where we can monitor the relevant quantities as a function of carrier concentration. A particularly interesting question to ask is whether (and if yes, to what extent) the physics of the intrinsic crystal at zero temperature is relevant to understanding the behavior of \Zstatcub{} and $C_Q$ in the doped regime. The choice of \sto{} is motivated by its popularity in perovskite oxide heterostructures, and of the availability of reference data for the \Zdyncub{}. Also, the transition to a negative electronic compressibility regime that was observed experimentally at LaAlO$_3$/\sto{} interfaces \cite{Tinkl2012} makes the study of the quantum capacitance in \sto{} especially worthwhile.

\begin{table}
\caption{Same as Table~\ref{tab:loo_becs} for \sto{}$(-0.05)$. The additional positive charge to compensate the doping is attributed to the Sr sublattice.\label{tab:sto_becs}}
\begin{ruledtabular}
    \begin{tabular}{c|cccc} 
    &Sr&Ti &Equat. O&Axial O\\\hline
pc & $4.102$&$7.615$&$-1.323$&$-3.911$\\
loc& $3.052$&$6.987$&$-1.473$&$-4.061$\\
\hline
diff. &1.050 & 0.628 & 0.150 & 0.150 \\
$\Delta Z^{\text{stat}}$ & 1.050 & 0.627 & 0.150 & 0.150 \\
\end{tabular}
\end{ruledtabular}
\end{table}

In Table~\ref{tab:sto_becs} we report the calculated values for \Zstatcub{} for \sto{}$(-0.05)$ with the two reference potentials. We see the same behavior as \loo{}, i.e., significant changes in the values when the reference potential is changed, predicted accurately by Eq.~(\ref{eq:ref_dep}). The values for the different sublattices roughly follow the undoped BECs, which we calculate to be 2.56, 7.29, $-2.06$, and $-5.73$ for Sr, Ti, the equatorial O, and the axial O, respectively. However, it should be noted that for the doping level, the changes from the undoped values are much more significant than \Zdyncub{} \cite{Dreyer2022}. 
For increasing doping, \Zstatcub{} tend to become more positive than their insulating counterparts, which can be directly linked to our discussion of the \loo{} case (see Sec.~\ref{sec:loo}).

\begin{figure}
	\includegraphics[width=\linewidth]{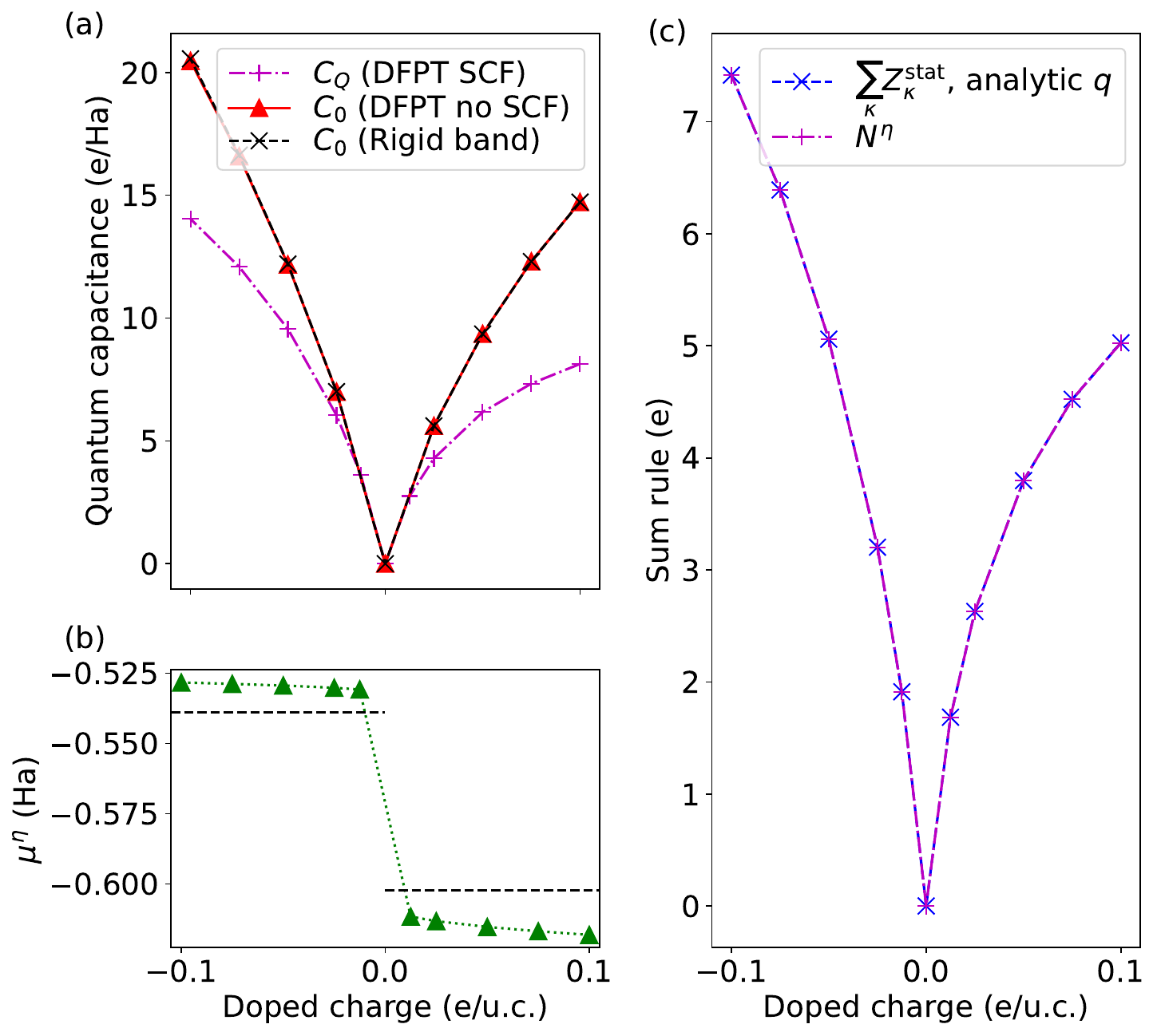}
	\caption{(a) Quantum capacitance for doped \sto{}, where positive (negative) charge corresponds to added holes (electrons). Calculated via DFPT with a scalar potential perturbation, both with and without self consistent fields (SCF), and via the ``rigid-band'' approximation, where the bands are fixed to the undoped case and the Fermi level adjusted. (b) Fermi level deformation potential versus doping. Dashed lines are the CBM and VBM deformation potentials for insulating \sto{}. (c) Quantities in the sum rule of \eq{eq:sum_rule2} for doped \sto{}. Charge response calculations were performed via the analytical long-wavelength expansion. All calculations are using the H+pc reference potential. }
	\label{fig:sto_chg}
\end{figure}

Finally, we summarize the results for the quantum capacitance and related quantities as a function of doping in Fig.~\ref{fig:sto_chg} [see also Table~\ref{tab:sum_rule} for \sto{}$(-0.05)$]. In panel (a), we plot the quantum capacitance with respect to doping, calculated either with or without SCF fields.
We see that the DFPT results without SCF match almost identically the ``rigid band'' value for $C_0$, i.e., Eq.~(\ref{eq:C_lind}). 
At low doping, $C_0$ and $C_Q$ are quite similar, and then increasing deviate as the free electron/hole density increases and electron-electron Coulomb repulsion plays more of a role. The asymmetry is indicative of the higher density of states of the conduction band, which is made out of Ti $3d$ states, versus the slightly more dispersive O $2p$ valence band.

In Fig.~\ref{fig:sto_chg}(b) we plot $\mu^\eta$ versus doping. Remarkably, this quantity has a rather weak dependence on the doping, apart from a discontinuity at zero as the crystal transitions from $p$-doped to the $n$-doped regime.
In the low-density limit, we expect that the intersect between $\mu^\eta$ and the $n=0$ axis to coincide with the relative deformation potential of the valence band maximum (VBM) and conduction band minimum (CBM), respectively. (Relative means that both are referred to the chosen potential reference, which follows either the pc or the loc conventions in our calculations.)
To test this idea, we have calculated the VBM and CBM relative deformation potentials in the insulating STO crystal; the results are shown as dashed lines in Fig.~\ref{fig:sto_chg}(b). The values are in good agreement with earlier calculations \cite{Stengel2015}, and closely match with the calculated values of $\mu^\eta$, thus proving our point. 
This analysis provides the numerical demonstration of the formal link between the \Zstatcub{} and the relative deformation potentials, which we established in Eq.~\eqref{eq:defpot}.

Finally, in Fig.~\ref{fig:sto_chg}(c) we demonstrate that the sum rule of Eq.~(\ref{eq:sum_rule2}) is accurately satisfied for all dopings. Note that here we use the analytic $q$ derivation method discussed in Sec.~\ref{sec:analytic_q}. Since $\mu^\eta$ depends weakly on doping, the $N^\eta$ (and therefore the sublattice sum of \Zstatcub{}) follow the same trend with doping as $C_Q$ discussed above.

\subsection{All-electron corrections}

Before closing, it's worth illustrating the dependence of the \Zstatcub{} on the pseudopotential approximation, by estimating the all-electron (AE) corrections discussed in Sec.~\ref{sec:ref_dep}. 
By applying Eq.~\eqref{eq_delta_z_2} to the AE core that we extracted from the atomic pseudopotential code, we obtain an effective core octupole of $\Delta O_{\rm Al}=-3.783$ $e\cdot$bohr$^2$ and $\Delta O_{\rm Cu}=-5.633$ $e\cdot$bohr$^2$ for the two free-electron metals of Sec. \ref{sec_el_metals}. 
(With the same method, we obtain $\Delta O_{\rm Mg}=-4.92$ $e\cdot$bohr$^2$, in excellent agreement with the value of $-4.85$ $e\cdot$bohr$^2$ reported by Hong and Vanderbilt.~\cite{Hong2011}) 
This, in turn, entails a correction to the static charge of, respectively, of $\Delta Z_{\rm Al}=1.23 e$ and $\Delta Z_{\rm Cu}=1.29 e$. (The larger octupole of Cu is compensated for by its much smaller quantum capacitance.)
These corrections are significant, and can become even larger in heavier elements. 
To demonstrate this point, we have also calculated the correction to the Os static charge in LiOsO$_3$, 
where we find $\Delta O_{\rm Os}=-22.41$ $e\cdot$bohr$^2$ and $\Delta Z_{\rm Os}=2.83 e$.

The fact that the \Zstatcub{} acquire an important contribution 
from Eq.~(\ref{eq:delta_z}) is worrisome: The main purpose of these quantities lies in their ability to capture the electron-phonon matrix elements in a certain class of materials \cite{Marchese2023}, and the latter shouldn't be affected by the inert core orbitals that move rigidly with their corresponding atom.
In our view, this result further questions the physical significance of \Zstatcub{} as bulk properties of the crystal.

\section{Summary and Conclusions \label{sec:conc}}

We have presented a critical analysis of the so-called static Born charges in metals, with special focus on their fundamental significance and physical interpretation. 
In the process, we have established an exact sum rule, which links the 
\Zstatcub{} to the charge response to a strain at fixed Fermi level. We have used this result to emphasize the intimate relation between the \Zstatcub{} and deformation potential theory, and the relevance of the self-consistent (i.e., including local Hartree fields and XC contributions) quantum capacitance of the crystal in the physics of static charge screening.
We have also discussed the dependence of these quantities on a number of approximations and conventions that are routinely used in practical implementations of the method, e.g., the length scale of the macroscopic averaging, the choice of the electrostatic potential reference and the details of the pseudopotentials.
The sensitivity of \Zstatcub{} to the technical details of the numerical implementation suggest that, though a useful quantity for, e.g., interpolation of interatomic force constants, \Zstatcub{} is unlikely to be an experimentally measurable bulk property of the crystal. 

To test these formal results, we have generalized density-functional perturbation theory by implementing the linear response to a uniform scalar potential, which we incorporated in the existing long-wave module of {\sc abinit}. This way, we could perform one-shot calculations of both the quantum capacitance and \Zstatcub{} directly at the $\Gamma$ point, by means of an analytical long-wavelength expansion and the $2n+1$ theorem.
Our results on elemental metals (Cu and Al), a compound metal (\loo{}), and a doped semiconductor (\sto{})
provide a stringent numerical benchmark for our main conclusions, and provide interesting insight on the physics of these materials.

As an outlook, our study motivates further investigation on the determination of the electronic currents in the lattice dynamics of dirty metals.
In this context, we regard the downfolding approach proposed 
in Ref. \cite{Berges2023} as a promising conceptual framework to treat the problematic Fermi-surface contributions, and understand how they are affected by scattering.
Similar ideas could help in dealing with static screening of polar phonons in lightly doped semiconductors \cite{Macheda2022}.
Finally, it would be interesting to clarify the role of dimensionality in the physical significance of both \Zstat{} and the quantum capacitance. 
We speculate, for example, that in a two-dimensional crystals many of the formal issues discussed in this work may not apply: the vacuum level acts as an obvious reference potential, and thereby should lift the main sources of arbitrariness that we have pointed out in 3D. (And indeed, the quantum capacitance of confined electron gases has been measured with high accuracy, see e.g., Ref.~\cite{Junquera2019} and references therein.)
We regard all these as a promising avenues for future study.

\begin{acknowledgments}
MS and AZ acknowledge support from the State Investigation Agency through the Severo Ochoa Programme for Centres of Excellence in R\&D (CEX2023-001263-S), from the Ministry of Science, Innovation and Universities (Grant No. PID2023-152710NB-I00) and from Generalitat de Catalunya (Grant No. 2021 SGR 01519). 
AZ thanks the CCQ at the Flatiron Institute for hospitality during 
the early stages of this work.
CED acknowledges support from NSF Grant No.~DMR-2237674. The Flatiron Institute is a division of the Simons Foundation.
We are grateful to Francesco Mauri, Francesco Macheda and Guglielmo Marchese
for illuminating discussions and a careful read of the manuscript.
\end{acknowledgments}

\appendix

\section{Geometric contribution to the quantum capacitance \label{app:int_pol}}

We can determine the ``geometric'' contribution to the quantum capacitance coming from the range separation approach by first isolating the $L$ dependence of $\chi^{\text{SR}}$. Suppose that the polarizability is nonsingular at a given ${\bf q}$;
then we have from Eq.~(\ref{eq:chi_SR_chi})
\begin{equation}
\label{eq:chi_inv_app_1}
\begin{split}
\chi^{-1} =& \epsilon_{\text{LR}} (\chi^{\rm SR})^{-1} \\
              =& (1 - v^{\text{LR}}\chi^{\rm SR}) (\chi^{\rm SR})^{-1} \\
              =& (\chi^{\rm SR})^{-1} - v^{\text{LR}}.
\end{split}
\end{equation}
Both terms in Eq.~(\ref{eq:chi_inv_app_1}) depend on $L$, though, as discussed in Sec.~\ref{sec:screened}, this dependence cancels to all orders. We can write an alternative decomposition as
\begin{equation}
    \label{eq:chi_inv_app_2}
\chi^{-1} = \tilde{\chi}^{-1} - v,
\end{equation}
where $\tilde{\chi}$ is given by
\begin{equation}
\tilde{\chi} = (1 + v \chi)^{-1} \chi 
\end{equation}
which is both an analytic function of ${\bf q}$ and $L$-independent. For the capacitance, we desire the $q\rightarrow 0$ limit; in that case, $\tilde{\chi}$ coincides exactly 
with the susceptibility function that one calculates within the PCM presciption \cite{Pick1970}. Comparing Eqs.~(\ref{eq:chi_inv_app_1}) and (\ref{eq:chi_inv_app_2}) gives
\begin{equation}
\tilde{\chi}^{-1} = (\chi^{\rm SR})^{-1} + v - v^{\text{LR}}= (\chi^{\rm SR})^{-1}+v^{\text{SR}},
\end{equation}
hence
\begin{equation}
(\chi^{\rm SR})^{-1} = \tilde{\chi}^{-1}-v^{\text{SR}},
\end{equation}
i.e., we have isolated all the $L$-dependence in $v^{\text{SR}}$. Using the  exponential Fourier filter of Eq.~(\ref{Eq_f_gaussian}), we have
\begin{equation}
     v^{\text{SR}}(q\rightarrow 0) = \frac{4\pi}{q^2}\left[1- \exp\left({-\frac{q^2 L^2}{4}}\right) \right]\simeq \pi L^2.
\end{equation}
Finally, using Eq.~(\ref{eq:chi_rho_met_a}) and the notation of Eq.~(\ref{eq:C_Hxc}), 
\begin{equation}
\begin{split}
     C_Q^{-1}&=-[\Omega\chi^{\text{SR}}(q=0)]^{-1}\\
     &=-(\Omega\tilde{\chi})^{-1}+\frac{v^{\text{SR}}}{\Omega} \\
     &=C_0^{-1}+\bar{C}_{\text{Hxc}}^{-1}+\frac{\pi L^2}{\Omega}.
\end{split}
\end{equation}

\begin{widetext}

\section{Proof of Eq.~(\ref{eq:delta_n}) \label{app:delta_n}}
Following Ref.~\citenum{Schiaffino2019}, we can generalize the geometric term for metals (in the adiabatic regime) as 
\begin{equation}
\begin{split}
\Delta n^{\alpha}_{\bf q}({\bf r}) =&  i \int [d^3 k] \sum_{m,n=1}^M (f_{m\bf k+q}-f_{n\bf k}) 
\langle u_{n \bf k}|{\bf r} \rangle \langle {\bf r}| u_{m \bf k+q} \rangle 
\langle u_{m \bf k+q}| \left( \hat{p}_{\bf k \alpha} + \frac{q_\alpha}{2}  \right) |u_{n \bf k} \rangle,
\end{split}
\end{equation}
where $\hat{p}_{\mathbf{k}\beta}=-i\partial/\partial r_\beta +k_\beta$ is the 
canonical momentum operator.
Defining the ground-state density operator as $\hat{\mathcal{P}}_{\bf k}$, this becomes
\begin{equation}
\begin{split}
  \Delta n^{\alpha}_{\bf q}({\bf r})=& i \int [d^3 k] \left[  \langle {\bf r}| \hat{\mathcal{P}}_{\bf k+q} \left( \hat{p}_{\bf k \alpha} + \frac{q_\alpha}{2}  \right) |{\bf r} \rangle -  
  \langle {\bf r}|  \left( \hat{p}_{\bf k \alpha} + \frac{q_\alpha}{2}  \right)  \hat{\mathcal{P}}_{\bf k} |{\bf r} \rangle \right] \\
  =& i \int [d^3 k] \left[  \langle {\bf r}| \hat{\mathcal{P}}_{\bf k} 
  \left( \hat{p}_{\bf k \alpha} - \frac{q_\alpha}{2}  \right) |{\bf r} \rangle -  
  \langle {\bf r}|  \left( \hat{p}_{\bf k \alpha} + \frac{q_\alpha}{2}  \right) \hat{\mathcal{P}}_{\bf k} |{\bf r} \rangle \right] \\
  =& i \int [d^3 k] \langle {\bf r}| [ \hat{\mathcal{P}}_{\bf k},  
  \hat{p}_{\bf k \alpha} ] |{\bf r} \rangle 
   - i q_\alpha \int [d^3 k] \langle {\bf r}| \hat{\mathcal{P}}_{\bf k} |{\bf r} \rangle \\
  =&  -\frac{ \partial n^{(0)}({\bf r}) }{\partial r_\alpha} - i q_\alpha n^{(0)}({\bf r}),
\end{split}
\end{equation}
where in the second line we have translated the first term in \textbf{k} space. This result is identical to Eq.~(46) of Ref.~\citenum{Schiaffino2019}.
\end{widetext}

\section{Free electron gas \label{app:feg}}

As a reference, in the following we recap some useful formulas in the context of the (either free or interacting) 
uniform electron gas.
The volume of the Fermi sphere is given by
\begin{equation}
V_{\rm F} = \frac{4\pi}{3} k_{\rm F}^3,
\end{equation}
where $k_{\rm F}$ is the Fermi velocity. The Fermi energy is given by
\begin{equation}
E_{\rm F} = \frac{\hbar^2}{2m_e} k_{\rm F}^2.
\end{equation}
Consider now a free electron gas on a lattice, with a unit cell volume $\Omega$. The total number of electrons per
cell is given by the ratio of the Fermi volume and the Brillouin zone volume, times a factor of 2 that accounts for the spin degeneracy. We have
\begin{equation}
N = 2 V_{\rm F} \frac{\Omega}{(2\pi)^3}.
\end{equation}
The charge density is then given by
\begin{equation}
n = 2 \frac{V_{\rm F}}{(2\pi)^3} = \frac{k_{\rm F}^3}{3\pi^2}. 
\end{equation}
We can also write it as a function of the Fermi energy as
\begin{equation}
n = \frac{\sqrt{2m_e E_{\rm F}}^3}{3\pi^2 \hbar^3}.
\end{equation}
Finally, we obtain the Fermi energy as a function of density,
\begin{equation}
E_{\rm F} = \frac{\hbar^2}{2m_e} (3\pi^2 n)^{2/3}.
\end{equation}
We can also write the density of states at the Fermi level as
\begin{equation}
\frac{dn}{dE} = \frac{m_e}{\pi^2 \hbar^3} \sqrt{2 m_e E_{\rm F}}.
\end{equation}
It is useful to express this as a function of density, 
\begin{equation}
\frac{dn}{dE} = \frac{m_e}{\pi^2 \hbar^2} (3\pi^2 n)^{1/3}.
\end{equation}

We move now to the treatment of electron-electron interactions, which we discuss with an exact Kohn-Sham framework in mind. We can work under the assumption that, in the relevant range of densities, the XC contribution is dominated by exchange, and
neglect correlation.
The exchange energy per cell is given by
\begin{equation}
    E_x = -\frac{3}{4} \left( \frac{3}{\pi} \right)^{\frac{1}{3}} n^{\frac{4}{3}} \Omega.
\end{equation}
The exchange potential is then given by
\begin{equation}
 V_x = - \left( \frac{3}{\pi} \right)^{\frac{1}{3}}  n^{\frac{1}{3}} 
\end{equation}
Finally, the exchange contribution to the inverse capacitance is
\begin{equation}
 \Omega C_x^{-1}  = - \frac{1}{3} \left( \frac{3}{\pi} \right)^{\frac{1}{3}}  n^{-\frac{2}{3}}.
\end{equation}
 
As an example, we work out Al explicitly. The cell volume of FCC Al is
$\Omega=a_0^3/4 = 106.3$ bohr$^3$. We have 3 electrons per cell, which yields
$n=0.02822$ bohr$^{-3}$. Assuming atomic units and a unit effective mass,
we obtain 
a Fermi level of $E_{\rm F}=0.44356$ Ha, corresponding to 12.07 eV. 
By using a more accurate value for the effective 
mass, $m^*=1.06$, we obtain $E_F=11.4$ eV. 
This yields a non-SCF capacitance of $\Omega dn/dE = 10.146$, which becomes $10.75$ 
if the effective mass factor is taken into account. This value is in excellent agreement 
with our first-principles result of $C_0 = 10.715$.

Regarding the exchange term, in Al one obtains $\Omega C_x^{-1}=-3.541314$, 
so
\begin{equation}
C \simeq ( C_0^-1 + C_x^-1 )^{-1} = 16.76.
\end{equation}
This is again in excellent agreement with our first-principles results.


%

\end{document}